\documentclass[useAMS,usenatbib]{mn2e}
\usepackage{graphicx}
\usepackage{eufrak}
\usepackage{eucal}
\usepackage{txfonts}

\begin{document}

\title[Dust in the early Universe]{Dust in the early Universe: Evidence for non-stellar dust production or observational errors?}

\author[Lars Mattsson]{Lars Mattsson\thanks{E-mail: mattsson@dark-cosmology.dk}\\
Dark Cosmology Centre, Niels Bohr Institute, University of Copenhagen, Juliane Maries Vej 30, DK-2100, Copenhagen \O, Denmark, and\\
Dept. of Physics and Astronomy, Div. of Astronomy and Space Physics, Uppsala University, Box 516, SE-751 20 Uppsala, Sweden}

\date{}

\pagerange{\pageref{firstpage}--\pageref{lastpage}} \pubyear{2010}

\maketitle

\label{firstpage}

\begin{abstract}
Observations have revealed unexpectedly large amounts of dust in high-redshift galaxies and its origin is still much debated.
Valiante et al. (2009, MNRAS, 397, 1661) suggested the net stellar dust production of the quasar host galaxy SDSS J1148+5251 may be
sufficient to explain the large dust mass detected in this galaxy, albeit under some very special assumptions (e.g., 'closed box' evolution and a 
rather high gas mass). Here it is shown that since accretion of essentially pristine material may lower the efficiency of dust formation significantly, 
and the observationally derived dust-to-gas ratios {for} these high-redshift galaxies are remarkably high, stellar dust production is likely 
insufficient. A model including metallicity-dependent, {non-stellar} dust formation ('secondary dust') {is presented}. The required contribution
from this non-stellar dust component appears too large, however. If all observational constraints are to be met, the resultant dust-to-metals ratio
is close to unity, which means that almost all interstellar metals exist in the form dust. This is a very unlikely situation and suggests the large
dust-to-gas ratios at high-redshifts may be due to observational uncertainties and/or or incorrect calibration of conversion factors for gas and dust 
tracers.

\end{abstract}


\section{Introduction}
Observations of damped Lya systems (DLAs), quasars and gamma-ray bursts (GRBs) have shown that relatively large amounts of dust are present in
high-$z$ galaxies \citep[see, e.g.][]{Ledoux02,Bertoldi03, Robson04, Beelen06, Michalowski08, Michalowski10a,Michalowski10b}.
This raises questions about the origins of cosmic dust and its
effect on cosmological observations. Most of the dust present in the Milky Way today is probably produced in the envelopes of evolved (age
$>1$ Gyr), low-mass stars. Observations of large amounts of dust present in high-$z$ galaxies with ages of less than 1 Gyr, on the other hand, 
suggest that even though low-mass stars seem to be the dominant dust producers in the present universe, this might not always have been the 
case \citep{Bertoldi03,Maiolino04}. 

Core-collapse supernovae (SNe) arising from early generations of stars have been suggested as the main sources of dust in the early Universe 
\citep{Todini01,Nozawa03,Dwek07,Bianchi07}. Models of dust formation in core-collapse SNe \citep{Todini01,Nozawa08} predict large enough amounts
of dust to account for the dust seen at high $z$. Direct observational evidence for core-collapse SNe as a major dust producers is
however still inconclusive, even for the local Universe \citep{Meikle07}. Observational results seem to suggest the onset of dust
condensation in SN ejecta, although the SN contribution to total mass of cosmic dust appears to be minor in most cases 
\citep[see, e.g.,][]{Kotak06,Kotak09,Gall10thes}.
However, a few intersting cases where significant amounts of dust are detected do exist, such as Keplers supernova \citep{Gomez09, Morgan03b} and Cassiopeia A
\citep{Barlow10, Dunne09}. There is also evidence for a significant amount dust ($0.4 M_\odot$) surrounding Eta Carina \citep{Gomez10}.
Whether the dust present in the Kepler and Cas A SN remnants is produced by the SNe themselves or the dust is formed from metals in the surrounding
ISM is unclear. It is also uncertain if {the} dust {which} can be formed in the winds of massive stars (such as Eta Carina) survive the shock wave of the
subsequent SN explosion. However, it is not possible to rule out SNe as significant dust producers, although a recent analysis of the existing
observational constraints on dust productivity in SNe has shown that, on average, SNe seem to be converting a much smaller fraction of {their} metals
into dust (corresponding to a few times $10^{-4}$--$10^{-2} M_\odot$ of dust) compared to AGB stars \citep{Gall10thes}.

Assuming SNe as the only source of dust in the early Universe, \citet{Dwek07} suggest a dust yield of $1M_\odot$ of dust per SN to account for dust 
masses in high-$z$ quasars. \citet{Valiante09} used the SN dust yields including dust destruction by the revese shock computed by \citet{Bianchi07}
and the currently only existing sets of dust yields for {low and intermediate mass (LIM)} stars by \citet{Ferrarotti06} and \citet{Zhukovska08} in
combination with the star-formation history obtained by simulation of a hierarchical assembly in a canonical $\Lambda$CDM cosmology \citep{Li07}
for the host galaxy of the second most distant ($z = 6.42$) quasar ever detected: SDSS J1148+5251. This galaxy appears to contain an unexpectedly high
dust mass (given its redshift) which has raised doubts about whether stars alone can produce so much dust in such a short time.
\citet{Valiante09} argued that stellar sources can account for the dust detected by observations and show also that the contribution of AGB stars to dust
production in the early Universe may be far from negligible.

Recently, \citet{Pipino10} reached a conclusion similar to that of \citet{Valiante09}, although their model includes also nonstellar sources of dust.
However, the model by \citet{Valiante09} as well as that of \citet{Pipino10} assumes a gas mass that is much larger (about an order of magnitude)
than the observed mass of molecular gas of $\sim 10^{10} M_\odot$ \citep{Walter04}. A more gas rich galaxy will obviously be able to hold
a greater mass of dust, which led \citet{Gall10thes} to suggest that the gas content of many dust rich galaxies at high redshifts may be
greatly underestimated (their preferred models has a few times $10^{11} M_\odot$ of gas after 1 Gyr).

An intriguing discovery which has complicated the picture is that not all quasar host galaxies at high redshifts show significant dust components.
In fact, some of them may be essentially dust free \citep[see, e.g.,][]{Jiang06}. Quasars at high redshifts are powered by supermassive black holes 
with masses $>10^9 M_\odot$ \citep[see, e.g.,][]{Vestergaard04,Jiang06}, which may be part of the explanation to why some quasar hosts are very
dust rich while others are not. \citet{Jiang10} have shown that there is a correlation between the estimated size of the central black hole and
the derived dust masses. This finding may lend some support to a hypothesis presented by \citet{Elvis02}, that dust may form in significant quantities
in the outflows of quasars.

In this paper it will be shown that the stellar dust production required to explain the observed dust mass in SDSS J1148+5251 and in other high-$z$
galaxies is higher than what appears reasonable taking both theoretical and observational constraints into account. Is some kind of 'secondary' dust 
production, possibly associated with a quasar outflow, required to explain the detections of large dust masses? Or are the high dust-to-gas ratios
at high redshifts simply an artefact due to incorrect calibration of conversion factors for gas and dust tracers?

\section{Theory and Model}
\label{theory}
\subsection{Cosmology and time of galaxy formation}
\label{cosmology}
Adopting a canonical $\Lambda$CDM cosmology, a closed form expression for the expansion scale factor $S(t)$ can be obtained
\citep{Gron02},
\begin{equation}
\left[{S(t)\over S(t_0)}\right]^3 = \frac{\Omega_M}{\Omega_\Lambda}
                      \sinh^2 \left( \frac{3H_0 \Omega_\Lambda^{\,1/2}} {2}\, t \right),
\end{equation}
where $H_0$ is Hubble's constant, $\Omega_M$ is the fractional matter energy density today, and
$\Omega_\Lambda$ is the fractional dark energy density today. 
Combining this expression with $z + 1 = S(t_0)/S(t)$ gives the time-redshift relation
\begin{equation}
\label{cosmol}
t_{\rm a}(z) = {2\over 3}\frac{H_0^{-1}}{\Omega_\Lambda^{\,1/2}}
    \sinh^{-1}\left[ \left(\frac{\Omega_\Lambda}{\Omega_M}\right)^{\!1/2} (z + 1)^{-3/2} \right].
\end{equation}
Here, a standard scenario with $H_0 = 70$~km~s$^{-1}$~Mpc$^{-1}$, $\Omega_M = 0.3$ and $\Omega_\Lambda = 0.7$ is assumed.
Galaxy formation is assumed to start at $z=14$, which corresponds to $t_{\rm a} \approx 290$ Myr. This choice of redshift is somewhat arbitrary, but
the first galaxies are believed to have formed during the reionisation epoch, i.e., between $z = 20$ and $z = 6$. Galaxy formation at $z=14$ may
thus be representative for when the vast majority of galaxies formed in the early Universe.

According to the current cosmological paradigm (see Sect. \ref{cosmology}), galaxies are formed by mergers with other galaxies and/or infall of 
baryons at a certain rate $\dot{ M}_{\rm i}$. The time scale of infall (mass assembly) may be defined as
$\tau_{\rm i} = {M_{\rm g} / \dot{ M}_{\rm i}}$.
It is reasonable to assume the star-formation rate somehow reflects the rate of infall. In particular, it is especially convenient (mathematically)
to assume $\dot{ M}_{\rm i} = \alpha\dot{ M}_\star$, which is equivalent to assuming that the gas mass is constant with respect to time 
\citep{Larson72}. This simplistic model of the rate of infall will be used throughout this paper.

 \begin{figure*}
  \resizebox{\hsize}{!}{
  \includegraphics{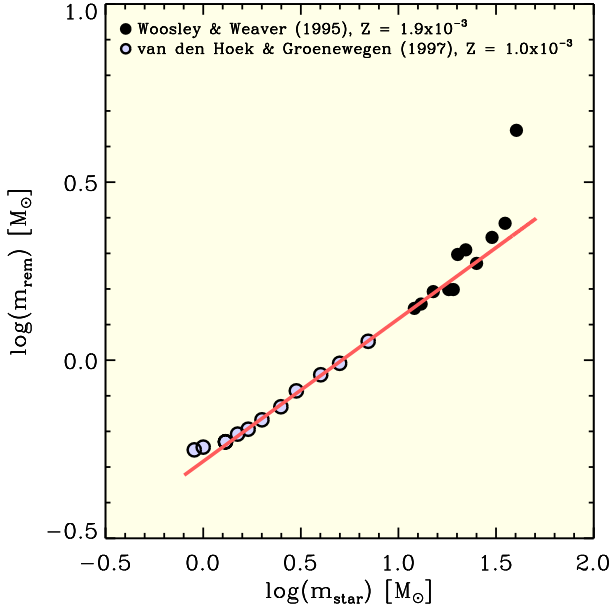}
  \includegraphics{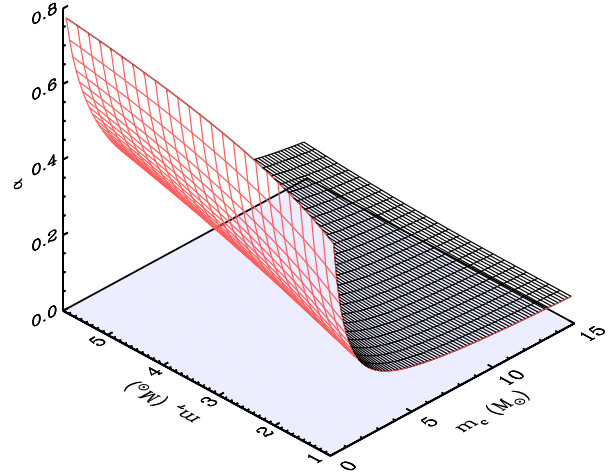}
  }
  \caption{Left: theoretical remnant masses as function of stellar mass compared to a simple power-law. Right: lock-up fraction as function of
           the turn-over mass $m_{\rm c}$ and the lower mass limit {$m_{\tau}$ for stars contributing to the build-up of dust and metals}.
           \label{remnants}}
  \end{figure*}

\subsection{Dust destruction}
\label{dustdep}
Dust in the ISM is expected to be destroyed by kinetic-energy injection to the ISM by SNe, but dust grains may also {grow} in the ISM, {in particular},
the outer parts of SN remnants. Further dust {growth} may also take place in molecular clouds. However, dust destruction {in the ISM} is probably dominating over
nucleation {(which rarely happens in the ISM)}, and it is thus fair to assume that introducing an {\em effective} rate of dust destruction in the ISM
$\dot{M}_{\rm ISM}$ is sufficient in order to model the dust cycle. Following \citet{Dwek07} the dust destruction time-scale is
\begin{equation}
\tau_{\rm d} = {M_{\rm g}\over m_{\rm ISM}\,R_{\rm SN}},
\end{equation}
where $M_{\rm g}$ is the gas mass, $m_{\rm ISM}$ is the effective gas mass cleared of dust by each SN event, and $R_{\rm SN}$
is the SN rate, which may be approximated as
\begin{equation}
\label{snr}
R_{\rm SN}(t) \approx \dot{ M}_\star(t)\int_{8M_\odot}^{m_{\rm u}} \phi(m)\,dm
\end{equation}
where $m_{\rm u}$ is the mass of the most massive SN-progenitors. The integral in Eq. (\ref{snr}) is a constant with repspect to time,
hence the time scale $\tau_{\rm d}$ may be expressed as
\begin{equation}
\label{taud}
\tau_{\rm d} \equiv {M_{\rm g}\over \dot{M}_{\rm ISM}} \approx \delta_{\rm ISM}^{-1} {M_{\rm g}\over \dot{ M}_\star},
\end{equation}
where $\delta_{\rm ISM}$ will be referred to as the efficiency of dust destruction in the ISM for each generation of stars. For a normal IMF and 
$m_{\rm ISM} \approx 1000 M_\odot$ \citep[the preferred value for the Milky Way, see][and references therein]{Dwek07}, this parameter is
$\delta_{\rm ISM} \approx 10.0$.

When considering the contribution from AGB stars there is a time-lag due to the lifetimes of these stars and therfore the destruction of dust depends on 
how the SN rate evolves. If the rate of star formation has decreased sufficienly when the release of new dust from AGB stars happens, the dust 
destruction by SNe will be much less due to the lower SN rate. Effectively the parameter $\delta_{\rm ISM}$ may have to be smaller than the number 
estimated above. \citet{Dwek07} considered $m_{\rm ISM}$ a free parameter, and found that with $m_{\rm ISM} = 100 M_\odot$
and a top-heavy IMF \citep[shallower than the][IMF]{Salpeter55} the estimated dust mass of SDSS J1148+5251 could be explained in terms of dust
production in SNe/high-mass stars. If $m_{\rm ISM} = 100 M_\odot$ is adopted as the effective gas mass cleared of dust by a SN-event, then
$\delta_{\rm ISM} \approx 1.0$.

Depletion of dust in the ISM need not be due to any actual destruction of dust. If a galactic wind is present it may have essentially the same
effect, although gas and atomic metals will be lost too. Since a galactic wind is beleived to be mainly driven by the kinetic-energy injection by SNe,
it is fair to assume that $\dot{ M}_{\rm w}$ is proportional to the star-formation rate $\dot{ M}_\star$ (see Eq. \ref{snr}). Hence, the outflow 
time-scale of the wind is
\begin{equation}
\tau_{\rm w} \equiv {M_{\rm g}\over \dot{M}_{\rm w}} \approx \delta_{\rm w}^{-1} {M_{\rm g}\over \dot{ M}_\star},
\end{equation}
where, by analogy with Eq. (\ref{taud}), {$\dot{M}_{\rm w}$ is the galactic-wind mass loss, and} $\delta_{\rm w}$ is the galactic 'wind efficiency'. However, 
for simplicity, only actual destruction of dust will be considered in the following, but the reader should bear in mind that a galactic wind may have a similar 
effect on the dust-to-gas ratio $Z_{\rm d}$.

\subsection{Equations of dust evolution}
\label{equations}
In order to obtain analytical solutions of adequate simplicity, the instantaneous recycling approximation \cite[IRA, which essentially means
all stars are assumed to have negligible lifetimes, see][]{Pagel97} is employed throughout the rest of this paper. No delayed element production due to 
stellar life-times is considered. This may, or may not, be a good assumption, but since the purpose of this paper is to provide 'proof of concept' and 
to study the dust component primarely during earliest phase of galaxy evolution, the IRA is a justified simplification.

A high-$z$ galaxy is assumed to be adequately modelled by a one-zone model and the production of metals have a negligible effect on
the evolution of the gas mass. Including infall of pristine gas and dust destruction in the ISM the eqautions for the
evolution of stars, gas and dust then becomes
\begin{equation}
{d M_{\rm s}\over dt} = \alpha \dot{M}_\star,
\end{equation}
\begin{equation}
{d M_{\rm g}\over dt} = -{d M_{\rm s}\over dt} + \dot{M}_{\rm i},
\end{equation}
\begin{equation}
{d M_{\rm d}\over dt} = y_{\rm d}{d M_{\rm s}\over dt} - {M_{\rm d}\over M_{\rm g}} \dot{M}_{\rm ISM}
\end{equation}
where  $\dot{ M}_\star$ is the star-formation rate $\alpha$ is the stellar lock-up fraction, i.e., the fraction (typically about 60-80\%) of the gas mass
converted into stars that will remain locked up in stellar remnants and very long-lived stars, $y_{\rm d}$ is the dust yield for a single generation of 
stars and $\dot{M}_{\rm i}$ is the infall rate. Combining the equations above and the prescriptions for infall and dust destruction one arrives at
\begin{equation}
\label{dem4}
M_{\rm g}{d Z_{\rm d}\over d t} = y_{\rm d}{d M_{\rm s}\over dt} - \left({\delta_{\rm ISM}\over \alpha} + \beta\right) {d M_{\rm s}\over dt}Z_{\rm d},
\end{equation}
where $Z_{\rm d} = M_{\rm d}/ M_{\rm g}$ is the dust-to-gas mass ratio and $\beta = 0,1$ for the closed-box and extreme infall cases, respectively.

\subsection{Lock-up fraction and stellar dust yields}
The primary\footnote{Dust is here assumed to form from newly synthesised elements.} dust yield $y_{\rm d}$ for a single generation of stars is defined as
\begin{equation}
\label{yield}
y_{\rm d} = {1\over \alpha} \int_{m_{\tau}}^{m_{\rm u}} p_{\rm d}(m)\,m\,\phi(m)\,dm.
\end{equation}
In Eq. (\ref{yield}) above, $p_{\rm d}$ is the fraction of the initial mass $m$ of a 
star ejected in the form of newly produced dust and $\phi(m)$ is the mass-normalised IMF. A convenient functional form of the IMF which includes
the turn-down at low stellar masses is the form suggested by \citet{Larson98},
\begin{equation}
\label{larson}
 \phi(m) = \phi_0\,m^{-(1+x)}\exp \left(-{m_{\rm c}\over m}\right),
\end{equation}
where $m_{\rm c}$ is the characteristic mass around which the turn-down begins and $\phi_0$ is a normalisation constant obtained from the condition
\begin{equation}
\int_{m_{\rm l}}^{m_{\rm u}} m\,\phi(m)\,dm = 1.
\end{equation}
The upper and lower mass-cuts, $m_{\rm u}$  and $m_{\rm l}$, respectively, represents the high-mass trunction and the brown-dwarf limit.

  \begin{table}
  \begin{center}
  \caption{\label{imfpar} IMF parameters and the resultant stellar lock-up fractions for the two considered IMFs.}
  \begin{tabular}{lllllll}
            & $x$  & $m_{\rm c}$ & $m_{\rm l}$ & $m_{\rm u}$ & $m_\tau$    & $\alpha$\\
            &      & $[M_\odot]$ & $[M_\odot]$ & $[M_\odot]$ & $[M_\odot]$ &        \\
  \hline
  Normal    & 1.35 & 0.35        & 0.1         & 100.0       & 3.0         & 0.63    \\
  Top-heavy & 1.35 & 10.0        & 0.1         & 100.0       & 3.0         & 0.10    \\
  \hline
  \end{tabular}
  \end{center}
  \end{table}

In the IRA the gas-return fraction $R$ of a generation of stars is obtained from the convolution of the IMF and the stellar remnant mass $w(m)$, i.e.,
\begin{equation}
R =  \int_{m_{\tau}}^{m_{\rm u}} [m-w(m)]\,\phi(m)\,dm.
\end{equation}
The remnant mass $w(m)$ is a function of the initial stellar mass $m$, which may be approximated by a simple power-law, $w(m) = k\, m^n$, where $k$ is
a constant. Using the remnant masses obtained by \citet{vandenHoek97} and \citet{WW95} for LIM and {high mass (HM)} stars, respectively, one obtains $k=0.5$ 
and $n=0.4$ for stellar masses less than $m = 35 M_\odot$ (see Fig. \ref{remnants}).

The lock-up fraction $\alpha$, used in the model equations discussed in Sect. \ref{equations}, is obtained as $\alpha = 1-R$. Adopting an IMF of
the form shown above in Eq. (\ref{larson}) and defining $G(a,b,c,z) \equiv \Gamma(a,z/c) - \Gamma(a,z/b)$, where
$\Gamma(a,z)$ is the incomplete gamma-function, one arrives at the expression (which holds for $x>1$, $n\le 1$)
\begin{equation}
R = {G(x', m_{\rm u}, m_\tau, m_{\rm c}) -
            {k \, m_{\rm c}^{n'}}\,G(x'-n', m_{\rm u}, m_\tau, m_{\rm c})
            \over G(x', m_{\rm u}, m_{\rm l}, m_{\rm c})},
\end{equation}
where $x' = x-1$, $n' = n-1$ and $m_\tau$ is the minimum stellar mass effectively contributing to the matter cycle at the considered redshift.
With the parameter values given in Table \ref{imfpar} and $m_\tau = 3.0 M_\odot$ (the initial mass for which the stellar lifetime corresponds to the
typical age of galaxies at $z = 5-6$) the lock-up fraction for a normal \citet{Larson98} IMF ($m_{\rm c} = 0.35 M_\odot$) becomes $\alpha = 0.63$, and
in the top-heavy ($m_{\rm c} = 10.0 M_\odot$) case $\alpha = 0.10$. As shown in the right panel of Fig. \ref{remnants}, $\alpha$ is a steep function
of the turn-over mass $m_{\rm c}$, but only weakly dependent on $m_\tau$. Hence, the choice of $m_\tau$ is not critical for either $\alpha$, nor the
effective primary yield $y_{\rm d}$.

\citet{Gall10thes} computed the total dust productivity of a generation of stars considering different IMFs and stellar dust-production efficiencies,
$\epsilon_{\rm d}(m) \equiv y_{\rm d}(m)/y_Z(m)$. Accordig to \citet{Gall10thes} the dust production efficiency derived from observations of SN
remnants is well-approximated by a power law, which connects nicely with theoretically expected efficiencies for AGB stars \citep{Ferrarotti06}. If this is a 
general law for stellar dust production, SNe are insignificant dust producers when weighted by an IMF. Theoretical work, as well as a handfull of 
{observational} studies, allows for a much higher efficiency in SNe. Using the results of \citet{Todini01}, \citet{Gall10thes} find that AGB stars 
would be insignificant relative to SNe as dust producers if no dust destruction {of SN-formed dust occur} due to the reverse shock.

The integrated (IMF-weighted) yields calculated by \citet{Gall10thes} is used in this paper, i.e., three cases are considered (see Table \ref{yields}): 
'observed' SN dust yields, theoretical SN yields with dust destruction due to reverse shocks \citep['high' yield,][]{Bianchi07}, and the theoretical 
'upper limit' {according} to \citet[][referred to as 'maximal' yield]{Todini01} -- all three in combination with the theoretical results for AGB 
stars \citep{Ferrarotti06,Zhukovska08}.

  \begin{table}
  \begin{center}
  \caption{\label{yields} Yields used in this paper.}
  \begin{tabular}{llll}
  IMF       & $y_{\rm d}^{obs}$   & $y_{\rm d}^{hi}$    & $y_{\rm d}^{max}$\\
  \hline
  Normal    & $8.81\cdot 10^{-4}$ & $1.52\cdot 10^{-3}$ & $1.08\cdot 10^{-2}$ \\
  Top-heavy & $3.51\cdot 10^{-3}$ & $1.55\cdot 10^{-2}$ & $1.88\cdot 10^{-1}$ \\
  \hline
  \end{tabular}
  \end{center}
  \end{table}

\subsection{Evolution of the dust-to-gas ratio}
\subsubsection{Closed-box evolution}
\label{closedbox}
A 'closed-box' evolution is obviously incompatible with the scenario described above in Sect. \ref{cosmology}. But for completenes, and since it corresponds to
the scenario studied by \citet{Valiante09} as well as \citet{Gall10b}, it is included also here. If there is no infall, the rate of change of
stellar mass is
\begin{equation}
\label{dem3}
{dM_{\rm s}\over dt} = \alpha\dot{M}_\star = -{dM_{\rm g}\over dt}.
\end{equation}
Assuming no dust destruction, combination of Eq. (\ref{dem3}) and Eq. (\ref{dem4}) gives the classical closed-box solution,
\begin{equation}
\label{simple}
Z_{\rm d} = y_{\rm d}\ln\left(1 + {M_{\rm s}\over M_{\rm g}}\right).
\end{equation}
Replacing $y_{\rm d}$ with the metal yield $y_Z$, this is also the solution for the metallicity $Z$ in a closed-box scenario.
If the dust-destruction term is included, the solution reads \citep{Edmunds01,Dwek07},
\begin{equation}
\label{modsimple}
Z_{\rm d} = {y_{\rm d}\over\nu}\left[1-\left(1+{M_{\rm s}\over M_{\rm g}}\right)^{-\nu}\right],\quad \nu = {\delta_{\rm ISM}\over \alpha}.
\end{equation}
This solution differs from the classical closed-box soultion in that $Z_{\rm d}\to y_{\rm d}/\nu$ as $M_{\rm s}/M_{\rm g} \to \infty$.
Assuming that dust grains may form and grow out of metals produced by previous generations of stars, i.e., a kind of 'secondary' dust, one may
redefine the dust yield as
\begin{equation}
\label{modyield}
\tilde{y}_{\rm d}(t) = y_{\rm d} + \epsilon\,Z(t),
\end{equation}
where $\epsilon$ is the fraction of the metals present in the ISM that will form dust grains. There is a natural upper limit to the value of $\epsilon$
since $Z_{\rm d} \le Z$, i.e., {\it the dust mass cannot exceed the total mass of metals available}.

With the modification above, the solution for $Z_{\rm d}$ becomes \citep[see also][]{Edmunds01}
\begin{equation}
\label{modsimple2}
Z_{\rm d} = {y_{\rm d}\over\nu}\left(1-\omega {y_Z\over y_{\rm d}}\right)\left[1-\left(1+{M_{\rm s}\over M_{\rm g}}\right)^{-\nu}\right]
          + \omega y_Z\ln\left(1 + {M_{\rm s}\over M_{\rm g}}\right)
\end{equation}
where $\omega \equiv \epsilon/\nu$. For the special case $\delta_{\rm ISM}=0$ (no dust destruction) the solution is
\begin{equation}
\label{modsimple3}
Z_{\rm d} = y_{\rm d}\ln\left(1 + {M_{\rm s}\over M_{\rm g}}\right) +
            {\epsilon \over 2}\,y_Z\left[\ln\left(1 + {M_{\rm s}\over M_{\rm g}}\right)\right]^2.
\end{equation}

\subsubsection{Galaxy formation by infall}
\label{infallmod}
Assume the rate of gas consumption is exactly balanced by the rate of infall.
With this assumption made, and no dust destruction term, the solution to Eq. (\ref{dem4}) is the 'extreme infall model' by \citet{Larson72}, i.e.,
\begin{equation}
\label{larson}
Z_{\rm d} = y_{\rm d}\left[1-\exp\left(-{M_{\rm s}\over M_{\rm g}}\right)\right].
\end{equation}
Adding dust destruction, the solution for $Z_{\rm d}$ has the same mathematical form as the \citet{Larson72} solution, but with an extra
parameter $\gamma$,
\begin{equation}
\label{modlarson}
Z_{\rm d} = {y_{\rm d}\over \gamma}\left[1-\exp\left(-\gamma{M_{\rm s}\over M_{\rm g}}\right)\right],\quad \gamma = \nu + 1,
\end{equation}
which reduces to Eq. (\ref{larson}) if $\delta_{\rm ISM}=0$ ($\gamma=1$).
Using the modified yield introduced in Eq. (\ref{modyield}) the solution for $Z_{\rm d}$ becomes
\begin{equation}
\label{modlarson2}
Z_{\rm d} = {y_{\rm d}\over \gamma}\left(1-\omega{y_Z\over y_{\rm d}} \right)\left[1-\exp\left(-\gamma{M_{\rm s}\over M_{\rm g}} \right)\right]
          + \omega y_Z\left[1-\exp\left(-{M_{\rm s}\over M_{\rm g}} \right)\right],
\end{equation}
where $\omega$ is defined as before, for the closed-box case.
For the special case $\delta_{\rm ISM}=0$ (no dust destruction) the solution is
\begin{equation}
\label{modlarson3}
Z_{\rm d} = y_{\rm d}\left[1-\exp\left(-{M_{\rm s}\over M_{\rm g}} \right)\right]
          + \epsilon y_Z \left[1-\exp\left(-{M_{\rm s}\over M_{\rm g}} \right)\left(1+{M_{\rm s}\over M_{\rm g}}\right)\right].
\end{equation}

\subsubsection{Evolved systems}
Even high-$z$ galaxies may have quite evolved stellar populations, e.g., they may be in (or approaching) a 'post-starburst state' after an intense
initial episode of {star formation}. One should note that the dust-to-gas ratio $Z_{\rm d}$ in the solutions above can never decrease (although $M_{\rm d}$
can) since $dM_{\rm s}/dt$ is always positive. Hence, it is of interest to know whether $Z_{\rm d}$ is approaching a constant value and thus has an 
upper limit or not.

For a closed-box scenario without dust destruction by SNe ($\delta_{\rm  ISM} = 0$), $Z_{\rm d}\to \infty$ as $M_{\rm s}/M_{\rm g}\to \infty$. Including
dust destruction ($\delta_{\rm  ISM} \ne 0$), there is an upper limit since then $Z_{\rm d}\to y_{\rm d}/\nu$ as $M_{\rm s}/M_{\rm g}\to \infty$, given
that $y_{\rm d}$ is treated as constant. Including 'secondary' dust production, $Z_{\rm d} \to \infty$ since the second terms of Eqs. (\ref{modsimple2})
and (\ref{modsimple3}) are simple functions the closed-box solution without dust destruction, which are monotonically increasing.

Any infall model where the rate of infall is proportional to the {star-formation} rate - not only the \citet{Larson72} model - has always an upper 
limit to $Z_{\rm d}$. In Larson's extreme infall model the metallicity can never exceed the yield. A similar upper limit exists also for a dust 
evolution model with infall, although the limit may be lower due to dust destruction. The dust-to-gas ratio will rapidly reach a constant value, i.e.,
$Z_{\rm d}\to y_{\rm d}$ as $M_{\rm s}/M_{\rm g}\to \infty$ if no dust destruction by SNe is present, and $Z_{\rm d}\to y_{\rm d}/\gamma$ if it is. With
the 'secondary' dust included, $Z_{\rm d}$ again approaches a constant value, this time given by
\begin{equation}
\label{evolved}
Z_{\rm d} \to \left\{
\begin{array}{lcc}
\displaystyle {y_{\rm d}\over \gamma} + \omega y_Z\left(1-{1\over \gamma}\right), &\delta_{\rm ISM} \ne 0, & \displaystyle{M_{\rm g}\over M_{\rm s}}\to 0,\\[4mm]
y_{\rm d} + \epsilon y_Z, & \delta_{\rm ISM} = 0, & \displaystyle{M_{\rm g}\over M_{\rm s}}\to 0.
\end{array}
\right.
\end{equation}

To sum up, $Z_{\rm d}$ has (within the present framework) an upper limit with only two exceptions:
\begin{enumerate}
\item a closed box without any dust destruction,
\item a closed box with 'seconday' dust production.
\end{enumerate}
Since a real galaxy cannot strictly be a 'closed box', it is fair to assume that there is an upper limit to the gas-to-dust ratio $Z_{\rm d}$ in most 
cases.

\subsection{Evolution of the dust-to-metals ratio}
A constant dust-to-metals ratio $\zeta\equiv Z_{\rm d}/Z$, is commonly adopted in models of galactic evolution \citep[e.g.][]{Edmunds98,Pei99}.
From observations, the dust-to-metals ratio $\zeta$ is known to be essentially constant in the local Universe \citep{Issa90}, while a lower (but similar) 
ratio is found at somewhat higher redshift \citep[see, e.g.][]{Vladilo98}. \citet{Edmunds98} considered a simplistic model where $\zeta$ was fixed, but 
it was later shown by \citet{Edmunds01} that $\zeta$ could, at least in principle, vary significantly over time. However, the $\zeta$-evolution 
approaches a constant value in many cases, so for evolved galaxies of similar type one may expect an approximately universal $\zeta$.

From Eqs. (\ref{modsimple}-\ref{modlarson3}) it is evident that the dust-to-gas ratio $Z_{\rm d}$ for the different models can be expressed as functions of the
metallicity $Z$. Hence, for a closed-box
\begin{equation}
\label{zeta1}
\zeta(Z) = \left\{
\begin{array}{lc}
\displaystyle\left({y_{\rm d}\over y_Z}-\omega\right)\left[1-\exp\left(-\nu {Z\over y_Z}\right)\right]\left(\nu{Z\over y_Z}\right)^{-1}+\omega, & \delta_{\rm ISM}\ne 0,\\[4mm]
\displaystyle{y_{\rm d}\over y_Z}+{\epsilon\over 2}{Z\over y_Z}, & \delta_{\rm ISM} = 0,
\end{array}
\right.
\end{equation}
and for the extreme infall case,
\begin{equation}
\label{zeta2}
\zeta(Z) = \left\{
\begin{array}{lc}
\displaystyle\left({y_{\rm d}\over y_Z}-\omega\right)\left[1-\left(1-{Z\over y_Z}\right)^\gamma\right]\left(\gamma{Z\over y_Z}\right)^{-1}+\omega, & \delta_{\rm ISM}\ne 0,\\[4mm]
\displaystyle{y_{\rm d}\over y_Z} +\epsilon\left[1+\left({y_Z\over Z}-1\right)\ln\left(1-{Z\over y_Z}\right)\right], & \delta_{\rm ISM} = 0.
\end{array}
\right.
\end{equation}
From the equations above it is obvious that the evolution of $\zeta$ leads to {different} 'end-states' depending on the assumptions made for each model.
$\zeta$ may thus reach a constant value, but can also grow without bound or approach zero as a galaxy evolves.

Since very high dust-to-metals ratios are not observed at any redshift
\citep[but see, e.g.][for examples of variations]{Galliano03,Vladilo04}, it seems the closed-box case with $\delta_{\rm ISM} = 0$ and 'secondary'
non-stellar dust production is less likely, or {star formation} has to essentially cease at
some point. It also appears that either the dust yield must be a significan fraction of the metal yield, or non-stellar 'secondary' dust production must
play an important role in the local as well as the high-$z$ Universe. {For} the Milky Way $\zeta_{\rm MW} = 0.3 - 0.5$, which is roughly the ratio
found in most local galaxies of similar type \citep[see, e.g.][]{Issa90}, and DLAs at $z = 0.7 - 2.8$ also show similar ratios \citep{Vladilo98}.

    \begin{figure*}
  \resizebox{\hsize}{!}{
  \includegraphics{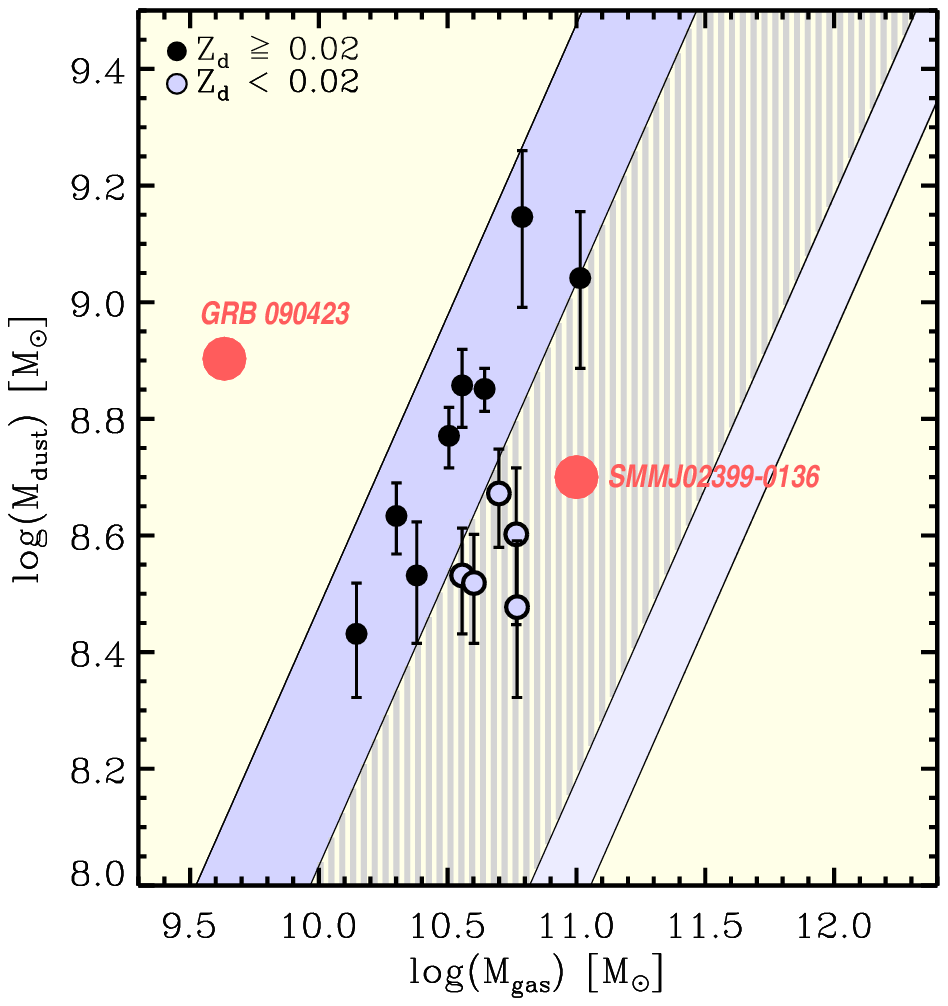}
  \includegraphics{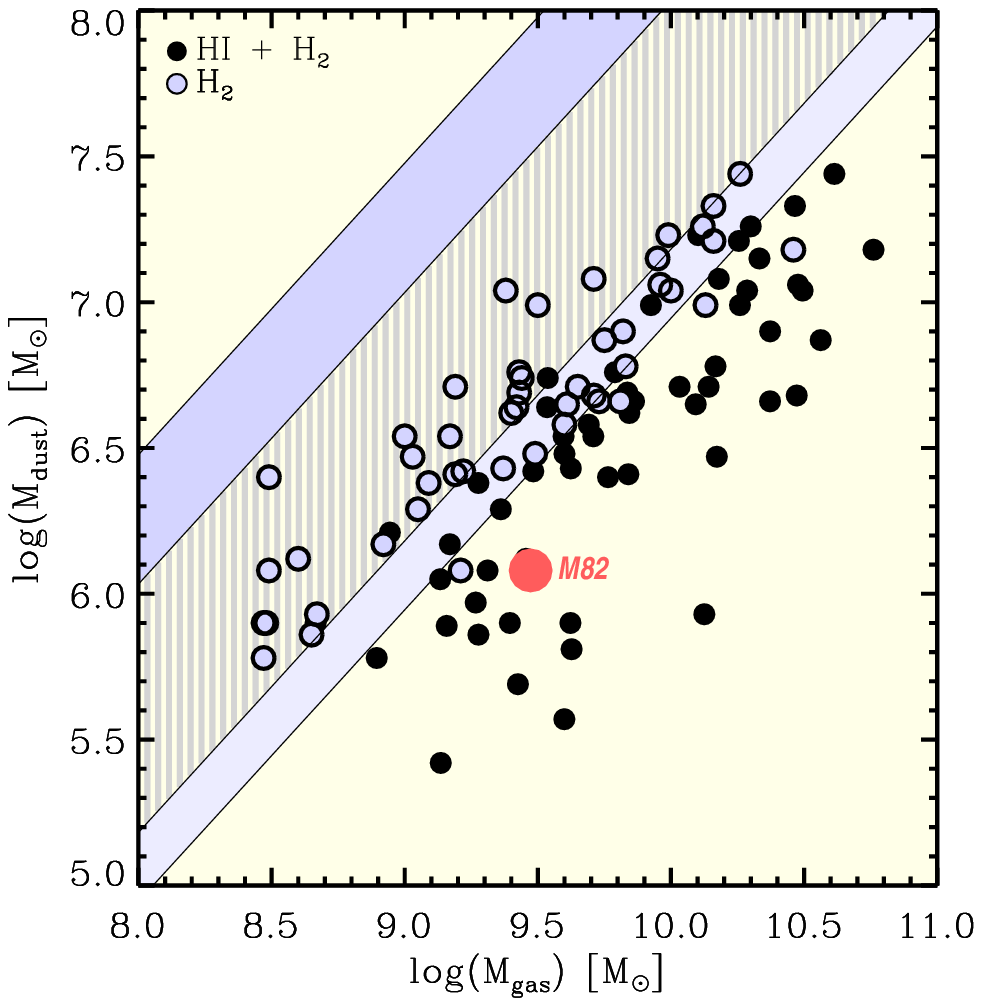}}
  \caption{Left: Comparison between evolved infall models with 'secondary' dust production (but without dust destruction) and observationally inferred dust masses
           of high-$z$ galaxies \citep[data taken from][]{Michalowski10a,Michalowski10b}. The light-blue region (appears light-gray in printed version) 
           is only accessable using the 'observed' yield. The darker blue/gray region can only be reached using the theoretical 'maximum' yield and the 
           hatched area in the middle corresponds to the region accessable using the 'high' yield. The plotted gas masses are twice the molecular hydrogen
           gas mass to account for neutral gas. For comparison the estimated upper limits of the dust and gas masses of the host/surroundings of
           GRB 090423 \citep[$z=8.2$,][]{Stanway10} are overplotted, as well as the derived dust and gas masses for the less distant sub-millimeter galaxy SMM 
           J02399-0136 \citep[$z=2.8$][]{Genzel03}.
           Right: Same models as in the left panel,
           but compared to local galaxies, with and without the inclusion of neutral hydrogen in the gas mass \citep[data taken from][]{Devereux90}.
           \label{qso}}
  \end{figure*}

\section{Results and Discussion}
\label{resudisc}
\citet{Valiante09} have suggested that stellar production of dust (in SNe and AGB stars) is high enough to explain the dust-mass estimates
of quasar host galaxies at high $z$, in particular the host galaxy of SDSS J1148+525. But is this true under all circumstances? \citet{Michalowski10b}
have shown that the minimal yield needed to explain the observationally inferred dust masses in sub-millimeter galaxies at $z>4$ is $15-65 M_\odot$ per
SN if the observational constraints on the stellar and gas mass are assumed to be correct. Their result is marginally consistent with the theoretical
maximum obtained by \citet{Todini01}, but does not account for any type of dust destruction.

The main result of the present paper is that only the most extreme, but also less likely, scenarios for cosmic dust production can reproduce the
dust masses inferred from observations of high-$z$ galaxies. That is a scenario which requires essentially no dust destruction - neither due to
the reverse shock in SN-explosions nor in the ISM due to kinetic energy injection by SNe - and extensive non-stellar dust production seems
necessary.

Another important result is that galaxy formation by merger events (and accretion of pristine gas), i.e., a cosmologically motivated rate of infall,
slows down the build-up of the interstellar dust component quite significantly. This {is} partly due to the trivial fact that the ISM is diluted 
from the unenriched gas falling into the galaxy, but also the fact that an infall scenario typically leads to significantly shorter dust destruction 
time scales (see Fig. \ref{taudplot}).

\subsection{Comparison with observed properties of high-$z$ galaxies}
It is reasonable to restrict a general comparison with observations to constraints derived for
the infall scenario, since in the current paradaign for galaxy formation galaxies are formed through mass assembly.
Furthermore, it can be assumed that dust rich high-$z$ galaxies are, despite their redshifts, in a relatively evolved state where a large
fraction of the baryonic mass is stars, i.e., the time scale for their assembly short.
A good starting point is therefore Eq. (\ref{evolved}) in Sect. \ref{infallmod}.

Using Eq. (\ref{evolved}) with $\delta_{\rm ISM} = 0$ one can obtain upper limits to the dust masses possible in an evolved system for a given dust yield and
efficiency $\epsilon$ of 'secondary' dust production. With $0\le \epsilon\le 1-y_{\rm d}/y_Z$ as the allowed range, Eq. (\ref{evolved}) constrains the possible
regions in the dust/gas-plane for given yields. In Fig. \ref{qso} evolved infall models with 'secondary' dust production (but without dust destruction) are
compared to observationally inferred dust masses of high-$z$ galaxies \citep{Michalowski10a,Michalowski10b} and local galaxies \citep{Devereux90}.

The light-blue region (grey in {printed version}) is only accessible using the 'observed' yield and the darker blue/grey region can only be
{reached} using the theoretical 'maximum' yield according to  \citet{Todini01}. The overlapping (hatched) region is accessible using either one of 
the 'high' and 'maximum' yield depending on the amount of 'secondary' dust production. The dust content of the galaxies with the highest dust-to-gas 
ratios can only be explained with the theoretical 'maximum' yield in combination with a relatively large $\epsilon$. This will be {discussed} in 
more detail below. Note that the gas mass for the high-$z$ galaxies are taken to be twice the observed H$_2$-mass to account for a neutral gas component, 
which cannot be detected in these objects \citep[the fraction of neutral gas has been estimated to be $\le 0.5$, see][]{McGreer11}.

 \begin{figure}
  \resizebox{\hsize}{!}{
  \includegraphics{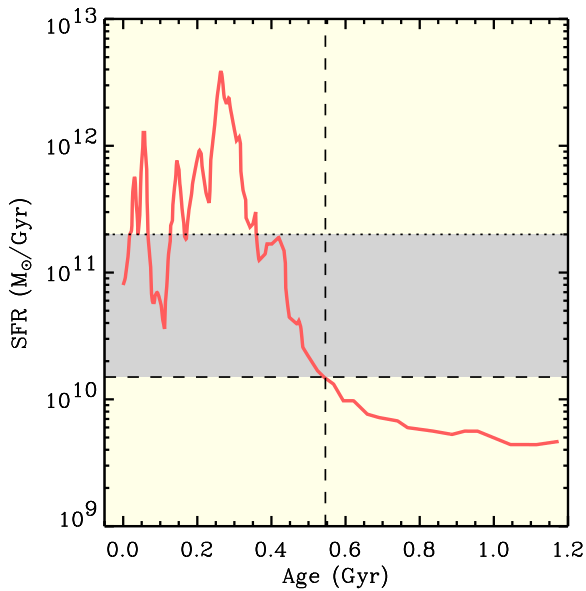}
  }
  \caption{Star-formation history in the host galaxy of SDSS J1148+5251 as predicted by the simulation of Li et al. (2007). The verical,
           dashed line marks $z=6.42$ and the horizontal, dotted line shows the star-formation rate predicted by
           the Kennicutt-Schmidt law adopting a total gas mass of $\sim~10^{10}M\odot$ (Walter et al. 2004).
           \label{sfr}}
  \end{figure}

As can be seen in Fig. \ref{qso}, the dust-to-gas ratios of high-$z$ galaxies are quite remarkable. Half of the sample considered here have 
$Z_{\rm d}> 0.02$, i.e., larger than
the total metallicity of the solar neighbourhood. One may argue that this fact implies gas masses which are about one order of magnitude larger than those 
obtained through observational estimates of the molecular gas \citep[see also models by][]{Gall10b}. However, the uncertainties associated with the gas
mass are certainly not on an order of magnitude scale. In fact, the total gas mass is not likely to be more than roughly a factor 2 larger if atomic
hydrogen and helium is added, and the uncertainty in the conversion factor between CO and H$_2$ ($\alpha$) can at most be a few times larger than the
value commonly used for high-$z$ objects. A value around $\alpha \sim 1$ is normally used for high-$z$ quasars \citep[see, e.g.][]{Downes98}, while
for local spiral galaxies $\alpha$, is typically taken to be $\sim 4.6$ \citep[see, e.g.][]{Solomon91}. For local starburst galaxies $\alpha \sim 1$, however.

It seems non-stellar dust production/growth may be required {to reproduce} the high dust-to-gas ratios. Dust may grow in dense molecular clouds {in} 
the ISM through accretion onto seed particles (most likely originating from stars) given enough time \citep{Draine90}. This component may be significant,
possibly a more important contribution than dust grown in stellar atmospheres and SN remnamts \citep{Zhukovska08,Asano10}. Furthermore,
\citet{Elvis02} have suggested dust may be {nucleated} (and grow) in broad emission line clouds in outflowing quasar winds. Both processes (growth in
the ISM and formation in quasar winds) can be indirectly associated with star formation. Stars form primarely in dense molecular clouds which is
also where dust grains in the ISM may grow through accretion. Quasar-wind {outflow rates} can be considered roughly proportional to the mass of the
supermassive black hole \citep[see][references therein]{Pipino10} and are thus linked to the {star formation} rate (or growth of the stellar mass)
assuming that the black hole is growing due to the galactic mass assembly \citep{DeBuhr10, Sarria10}. Hence, the non-stellar dust production/growth
may effectively be modelled by introducing a modified (metallicity-dependent) yield as in Eq. (\ref{modyield}). The possibility of grains
forming in quasar ourflows and grains growing in the ISM should be investigated further by more detailed modelling before one concludes that this
is the solution to the dust formation problem in the early Universe, however.

Although significant non-stellar dust production is an attractive idea, much of the dust production problem at high $z$ can be
resolved by assuming the true gas mass may be a few times higher due to a significant atomic component and uncertainties in the derivation of the
molecular component (see above), combined with the fact that dust masses are hard to quantify with high precision. It may very well be that the dust
masses derived for, e.g., high-$z$ quasars are too large due to the fact that the model for the dust emission is quite rudimentary. Dust mass
estimates are generally uncertain up to a factor of a few \citep{Silva98}. Hence, the true dust-to-gas ratio $Z_{\rm d}$ could, at least in principle,
be an order of magnitude lower, which {could then} be explained by stellar dust production {alone}. If $Z_{\rm d}$ is only a few times
$10^{-3}$, the required dust yield is reasonable from a theoretical point of view {(c.f. the effect of reverse shocks)} and consistent with results for
local galaxies (see Fig. \ref{qso}).

\subsection{SDSS J1148+525}
\label{SDSS}
The host galaxy of the SDSS J1148+5251 quasar has received considerable attention in the literature since it appears to have an unexpectedly large
dust mass for its high redshift ($z = 6.42$), which seems to require a rather dramatic early {star formation} history if dust is mainly produced in stars
\citep[see, e.g., the models by][]{Dwek07,Valiante09,Gall10b,Pipino10}. This quasar host galaxy is used as a test bench also in the present paper.
SDSS J1148+525 provides insight on the evolutionary states of dust-rich quasar hosts at high redshifts.

{Massive high-$z$} galaxies seem to have reached a relatively evolvled state, despite their high redshifts, indicating that they form in a rapid
series of mergers/infall events as in the model by \citet{Li07}. The time scale of the build-up of the stellar mass is clearly not more than a few Myr. As a
consequence, they have typically reached solar, or even supersolar, metallicities and the subsequent evolution (from the observed state until present
time) will neither change the metallicity, nor the dust-to-gas ratio dramatically.

The age of the Universe at $z=6.42$ in the {standard} $\Lambda$CDM model is $t_{\rm a} \approx 840$ Myr and galaxy formation is here assumed to start at 
$z=14$ corresponding to $t_{\rm a} \approx 290$ Myr. The host galaxy of SDSS J1148+5251 has in such case evolved for 550 Myr prior to its observed state.
\citet{Li07} constructed a merger tree, extracted from a cosmological simulation, to follow the hierarchical mass assembly of a galaxy thought to be similar to
the host galaxy of SDSS J1148+5251. Here, just as in the work by \citet{Valiante09}, their resultant star-formation history (SFH) is used as input (see
Fig. \ref{sfr}). In practice, however, the quantity used is the time-integrated star-formation rate, i.e., the stellar mass,
\begin{equation}
M_{\rm s}(t) = \alpha\int_0^t \dot{M}_\star(t')\,dt',\quad t = t_{\rm a}(z) - t_{\rm a}(z=14),
\end{equation}
which was computed numerically from the data shown in Fig. \ref{sfr} using a simple Romberg routine. In contrast to \citet{Valiante09}, a scaled-down version
of this SFH is considered to meet observational constraints on the total mass.

Using the analytic model described in Sect. \ref{equations}, 150 parameter configurations are considered (see Appendix \ref{grid}) in order to find the 
most likely models given some observational constraints to the dust mass in SDSS J1148+5251 \citep[see][and references therein]{Valiante09} and the 
dust-to-metals ratios found in damped Ly$\alpha$ absorbers at lower redshifts \citep[$1\le z \le 2.5$][]{Vladilo98}. Although inconsistent with a merger 
scenario, the closed-box case is here considered because (1) previous studies \citep{Valiante09,Gall10b} have considered the SDSS J1148+5251 quasar host 
galaxy to be a 'closed box', (2) it is important to analyse the net differeces between an infall and closed-box scenario, since the evolution of the
gas mass affects net dust-destruction efficiency (see Eq. \ref{taud}) and (3) infall always lowers the efficiency of metal production \citep{Edmunds90}.

  \begin{figure*}
  \resizebox{\hsize}{!}{
  \includegraphics{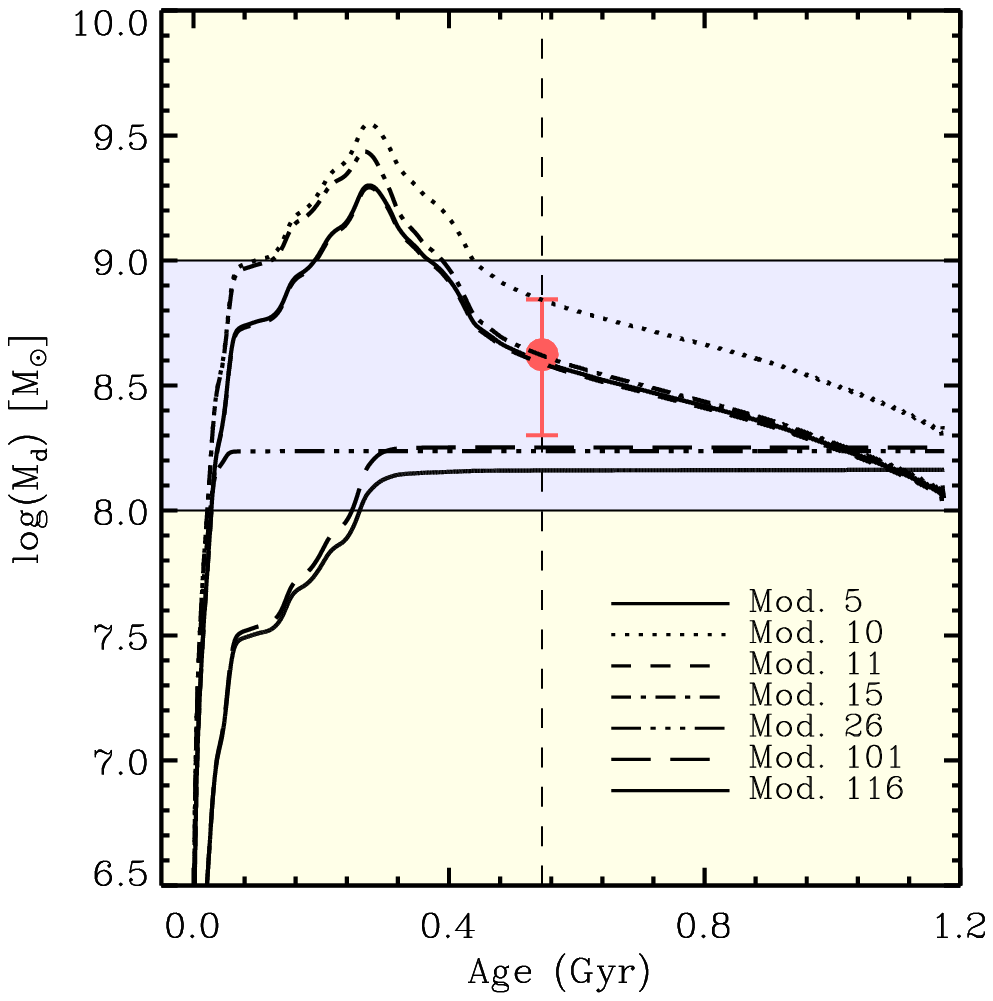}
  \includegraphics{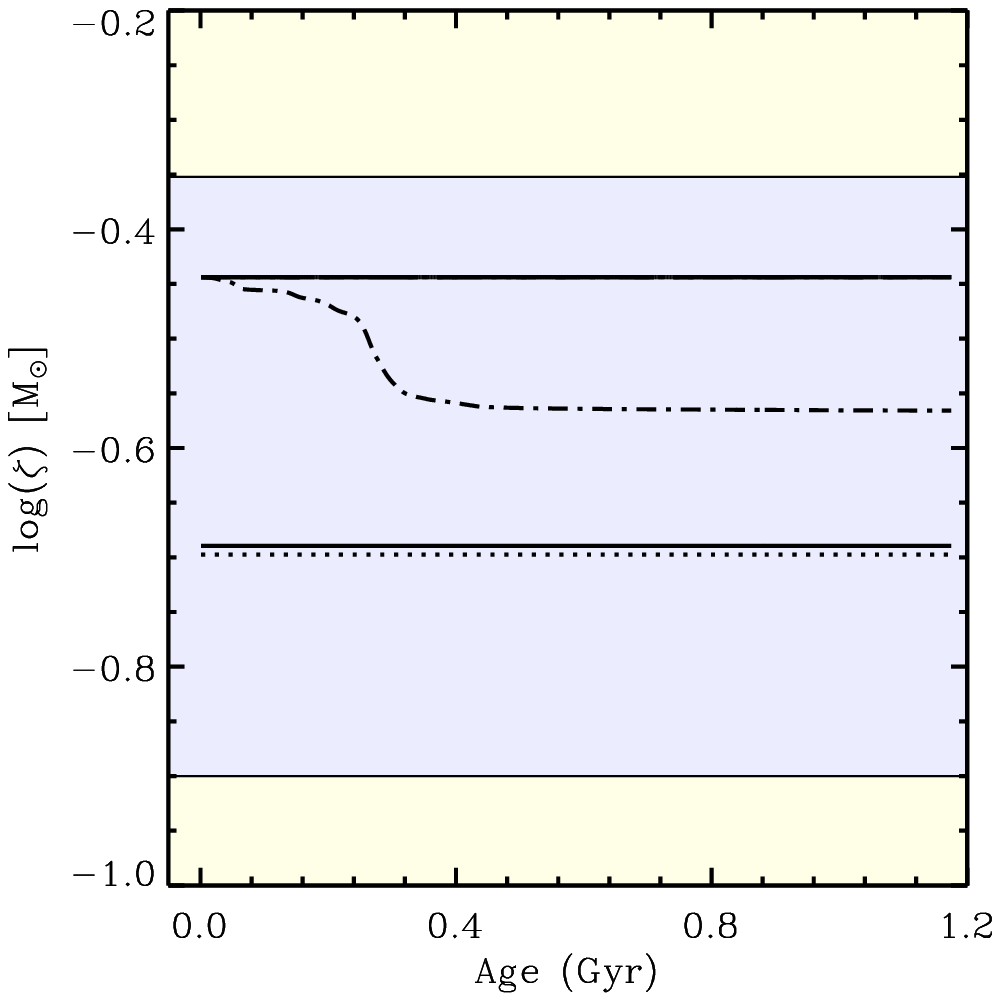}}
  \caption{Best-fit models for SDSS J1148+525. The left panel shows the evolution of the dust mass. The grey region indicate the range of dust masses considered to
  be consistent with observations. The red circle with error bars show the average observationally deduced dust mass. The right panel shows the dust-to-metals
  ratio for the same models. The grey region again indicate the expected range of dust-to-metals ratios based on observations of DLAs at moderately high redshift.
           \label{bestfit}}
  \end{figure*}

  \begin{table}
  \begin{center}
  \caption{\label{masscomp} {Baryonic} mass components and IMFs used to model SDSS J1148+525.}
  \begin{tabular}{lllll}
  Models       & $M_{\rm s}$ [$M_\odot$] & $M_{\rm g}$ [$M_\odot$] & $m_{\rm ISM}$ [$M_\odot$] & IMF\\
  \hline
  1-30         & $8.7\cdot 10^{11}$ & $1.6\cdot 10^{10}$ & 0, 100, 200, 1000 & Normal \\
  31-60        & $1.4\cdot 10^{11}$ & $1.6\cdot 10^{10}$ & 0, 20, 100, 200   & Top-heavy \\
  61-90        & $8.7\cdot 10^{11}$ & $4.3\cdot 10^{11}$ & 0, 100, 200, 1000 & Normal \\
  91-120       & $2.9\cdot 10^{10}$ & $1.6\cdot 10^{10}$ & 0, 100, 200, 1000 & Normal \\
  121-150      & $2.9\cdot 10^{10}$ & $1.6\cdot 10^{10}$ & 0, 20, 100, 200   & Top-heavy \\
  \hline
  \end{tabular}
  \end{center}
  \end{table}

Only a few of the 150 models seem to reproduce both the expected dust mass and a reasonable dust-to-metals ratio $\zeta$ (the models presented in Fig. 
\ref{bestfit}). A plausible dust-mass range for the SDSS J1148+525 host is that adopted by \citet{Valiante09}, i.e., $2-7\cdot 10^8 M_\odot$. Given the 
spread and uncertainty of the dust masses obtained by different authors {\citep[see][and references therein]{Valiante09}}, a dust mass in the range 
$10^{8-9} M_\odot$ is {here} considered to be consistent with the observations. A 'reasonable' range for the dust-to-metals ratio is here taken to 
be the same as for the sample of DLAs at medium high redshift studied by \citet{Vladilo98}, {as} $\zeta$ is not changing much after the intial phase 
of galaxy evolution unless the grain-growth rate is particularly low \citep[see, e.g.][Fig. 4]{Edmunds01}. Assuming dust of Galactic type, the DLAs 
considered by \citet{Vladilo98} places this quantity between 42\% and 89\% (typically 60\%) of the Galactic value, where the latter is 
$\zeta \approx 0.5$ \citep{Meyer98,Whittet91}.

The grid of models can be divided into five sub-sets of models based on the assumptions made regarding total mass, stellar mass, gas mass (at $z= 6.42$),
{dust destruction} and the IMF (see Tables \ref{imfpar} and \ref{masscomp}).
The first sub-set (models 1-30) represents an attempt to reprouce the metals and dust assuming the star formation builds up to roughly the same
stellar mass as in the simulation by \citet{Li07}, while still maintaining the lower-limit gas mass suggested by observations. With model 11 as the
only exception, non-stellar dust production appears necessary to reach {above $10^8 M_\odot$} of dust at $z = 6.42$. A few models (6, 12 and 26) 
{do} reach above $M_{\rm d} = 10^8 M_\odot$ and may be considered marginally consistent with observations in that regard. Model 26 is the
only infall model without non-stellar dust production (among models 1-30) which is consistent with the dust mass derived from observations (see Fig.
\ref{bestfit}). It is also consistent with the range of dust-to-metals ratios one would expect for a galaxy like SDSS J1148+525 (see Fig.
\ref{quasar_zeta}).

Changing the IMF from a normal \citet{Larson98} IMF ($m_{\rm c} = 0.3 M_\odot$) to at top-heavy ditto ($m_{\rm c} = 10 M_\odot$) increases both the amount
of dust production as well as the dust destruction by SN shocks, as can be seen from models 31-60. Models 36, 42, 51 and 57 are all reproducing the 
expected range for the dust mass, while all models with higher dust destruction efficiency ($m_{\rm ISM} > 20 M_\odot$) predict too much dust destruction 
due to the significantly increased SN-rate. The resultant dust masses are in those cases at most a few times $10^7 M_\odot$. Models 31-60 are essentially 
ruled out because the resultant metallicities are too high to be realistic ($Z = 1.47$ at $z = 6.42$) and inconsistent with the expected range {of}
dust-to-metals ratios (see Fig. \ref{quasar_zeta}).

Models 61-90 (see  Fig. \ref{quasar_zeta}, and Table~\ref{parameters_g}) are analogous to the numerical model by \citet{Valiante09} in that they 
have similar stellar and gas masses (a total baryon mass of $M_{\rm tot} = 1.3\cdot 10^{12}M_\odot$) and represent a closed-box scenario. The stellar 
dust yield referred to as the 'high' yield (see Table~\ref{yields}) and corresponds to the model preffered by \citet{Valiante09}. The resultant gas mass
($M_{\rm gas} = 4.3\cdot 10^{11} M_\odot$) is significantly higher than the mass inferred from observations \citep[see, e.g.][]{Walter04}.
With this high gas mass, even the low ('observed') dust yield \citep[see][]{Gall10b} is sufficient to reproduce the {observed dust mass}, and
the 'maximal' yield leads to an over-production of dust in several cases (models 71-73 and 86-88).
This is the main reason why the model by \citet{Valiante09} is marginally consistent with the high dust mass derived from observations without invoking
dust growth in the ISM or any other seconday dust source. More gas simply means more dust without resulting in unrealistic dust-to-gas
ratios. But as one can see in Fig. \ref{quasar_zeta}, the total amount of metals is generally too high to have {dust-to-metals ratios} in the expected range.

Applying observational constraints on both the gas mass and the stellar mass (models 91-120) results in two plausible models. Both are models without
dust destruction {in the ISM} and the 'maximal' stellar yield (see Table \ref{yields}). Adding non-stellar dust production (as it is implemeted here)
means no actual improvement for the models with lower yields. It essentially requires that all metals in the ISM are turned into dust.

Adding a top-heavy IMF (models 121-150, see Fig. \ref{quasar_sth}) leads to a result very similar to that of models 31-60 (also with a top-heavy IMF), i.e.,
the over-all metal production is not sufficient to reproduce the expected range of {dust to metals}.

\subsection{Infall vs. closed box}
Among the seven models of SDSS J1148+525 that reproduce the expected dust mass and dust-to-metals ratio (see previous section), only two are infall models.
One may note as well that all the closed-box models presented here \citep[and the model by][]{Valiante09} are conceptually inconsistent, since
\citet{Li07} consider a galaxy being formed by a sequence of merger events, where the smaller progenitor galaxies are very metal poor and have had very
little star formation ($M_{\rm g}/M_{\rm tot}\approx 1$). The infall models (16-30, 46-60, 76-90, 106-120 and 136-150) are therefore more appropriate.
This fact was indeed acknowledged by \citet{Valiante09}, although the exact consequences were not discussed. 

\begin{figure}
  \resizebox{\hsize}{!}{
  \includegraphics{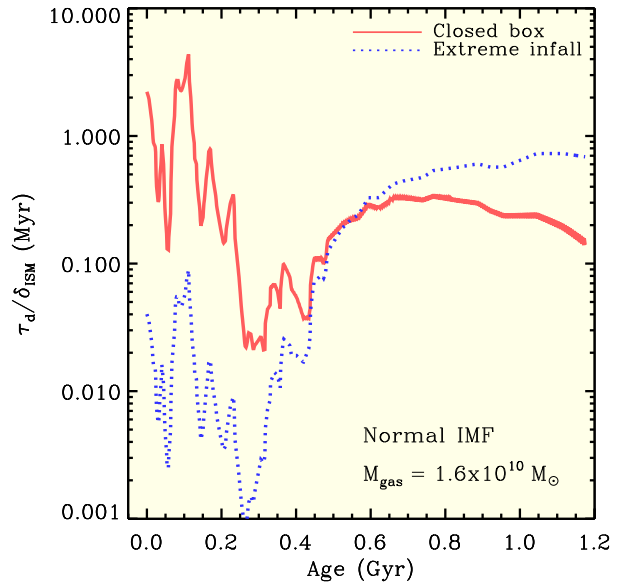}}
  \caption{Dust destruction time scale as function of time for a closed-box and an extreme infall scenario using the star-formation history
           resulting from the simulation of Li et al. (2007).
           \label{taudplot}}
  \end{figure}

From the present study it is clear, however,
that infall models are not preferred over closed-box models in that they {do} not reproduce the dust mass and/or overall metallicity better than
closed-box models. Infact, a merger/infall scenario requires a higher stellar dust yield in order to produce the same dust mass, which is due to the
effects of the dilution by pristine infall \citep[see, e.g.][]{Edmunds90} and the fact that the dust destruction time scale is proportional to gas mass.
The latter is important during the early phase of evolution, since in a closed-box scenario the initial gas mass is always equal to the total mass of
baryons. The gas mass is never as high as during the initial phase of the gas mass evolution, which means that the dust destruction time scale is
significantly longer at early times (when the rate of dust {production} is at its highest) compared to the infall scenario (see Eq. \ref{taud} and 
the example in Fig. \ref{taudplot}).

In the model of SDSS J1148+525, the intense star formation expected during the first 400 Myr would produce a lot of stellar dust, while $\tau_{\rm d}$
would still be relatively long on average in the closed-box case. As a consequence, the effective dust production {have to be} higher. If an infall 
scenario is adopted, the gas mass during the early evolution is lower even if the final total mass of baryons is the same. For example, in models 1-30 
the dust destruction time scale is about 15 times shorter for infall models compared to closed-box models and therefore infall models have a lower 
dust-formation efficiency.

Infall models always require a larger metal yield than the closed-box models in order to {contain} the same amount metals.
As discussed above, the dust-to-metals ratio $\zeta$ should be roughly that of DLAs \citep{Vladilo98}. The lower
net-production of dust for a given stellar mass of infall models tends to result in higher $\zeta$-values (see Fig. \ref{quasar_zeta}).
The infall models presented here are perhaps unrealistic in that they assume all gas ejected by evolved stars and all infalling gas is converted into new
stars (keeping $dM_{\rm g}/dt = 0$). But the dust yield $y_{\rm d}$ required to produce the estimated dust mass is at the high end and the
star-formation efficiency {will} under all circumstances be high in quasar host galaxies at high redshifts.

\subsection{Quasar outflows}
As argued in Sect. \ref{dustdep}, the {presence} of a galactic wind driven by SNe will have roughly the same effect on the dust mass as dust destruction
by SNe has, thus making it even harder to obtain the high dust masses in some high-$z$ galaxies. However, the outflows associated with quasars may also 
have a totally different effect. \citet{Elvis02} suggested that dust may be created in broad emission line clouds associated with quasar outflows. As
opposed to the gas in the ISM, these gas clouds could in principle (remains to be proven) have both nucleation and growth of dust grains. This fact
makes quasar outflows a possible source for the 'missing dust' in models of dust formation in high-$z$ quasar host galaxies. The models presented here
suggest the efficiency of dust production must be quite high, i.e., a large fraction of the atomic metals in the gas must be converted into dust
(essentially all the metals in the galaxy need to be in dust in many cases, see tables in Appendix \ref{grid}). This might be a problem, since one can
only have as much metals as can be produced by the stars and the total gas mass in a quasar outflow is not likely to be very
large. Thus quasar outflows may contribute, but the contribution can be considered negligible \citep[see also modelling results by][]{Pipino10}.

\subsection{A top-heavy IMF?}
There is a variety of evidence for top-heavy IMFs in starburst environments \citep[see, e.g.,][]{Dabringhausen09}.
\citet{Valiante09} showed that in their model LIM stars contributed about 80\% of the dust, using {a} \citet{Larson98} IMF, which is similiar to that of
\citet{Salpeter55} except at low masses where the \citet{Larson98} IMF has a smooth turn-over. But with a top-heavy IMF the situation is quite different.

From observations it seems LIM stars are producing significantly more dust per stellar mass than HM stars (cf. the 'observed' yield in Table 
\ref{yields}). Hence, a top-heavy IMF would under these circumstances result in a lower total stellar dust yield $y_{\rm d}$. If the dust production is
dominated by HM stars, on the other hand, a top-heavy IMF leads to a significantly increased dust production \citep{Gall10b}. {However,} a larger
fraction of HM stars also increases the SN-rate, which in turn leads to more interstellar dust destruction due to SN-shocks (see Sect. \ref{SDSS},
Fig. \ref{quasar_zeta}, Models 31-60 and 121-150 in Appendix \ref{grid}). But if this type of dust destruction {is} negligible, an initially 
top-heavy IMF may help to boost dust production in the early Universe. In general, {though}, a top-heavy IMF means no particular improvment since 
(1) dust destruction due SN-shocks is expected and (2) the dust-to-metals ratio changes in an unfavourable direction (again, see Fig. \ref{quasar_zeta},
{where it is clear that the ratios are too low}).

  \begin{figure*}
  \resizebox{\hsize}{!}{
  \includegraphics{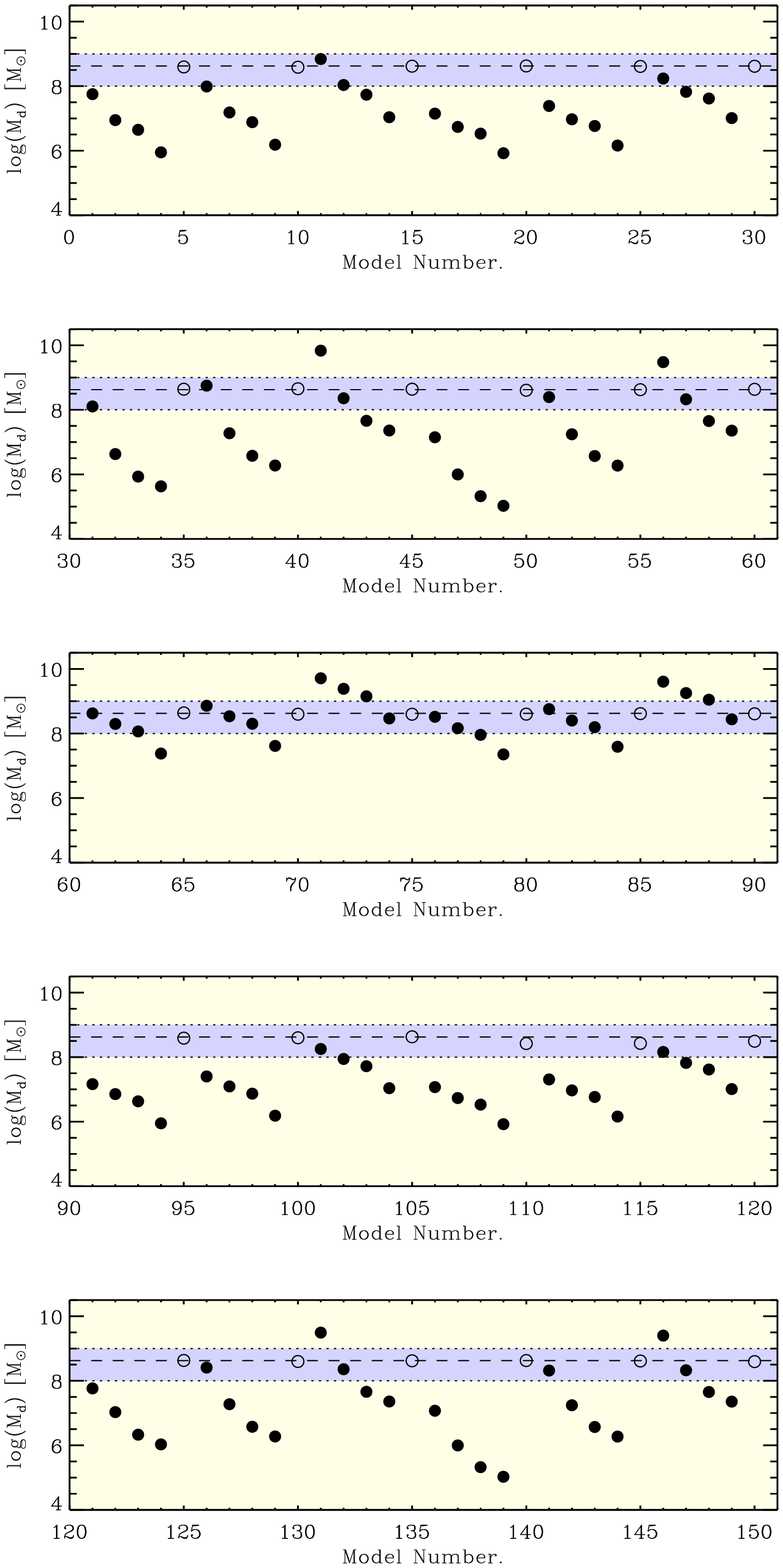}
  \includegraphics{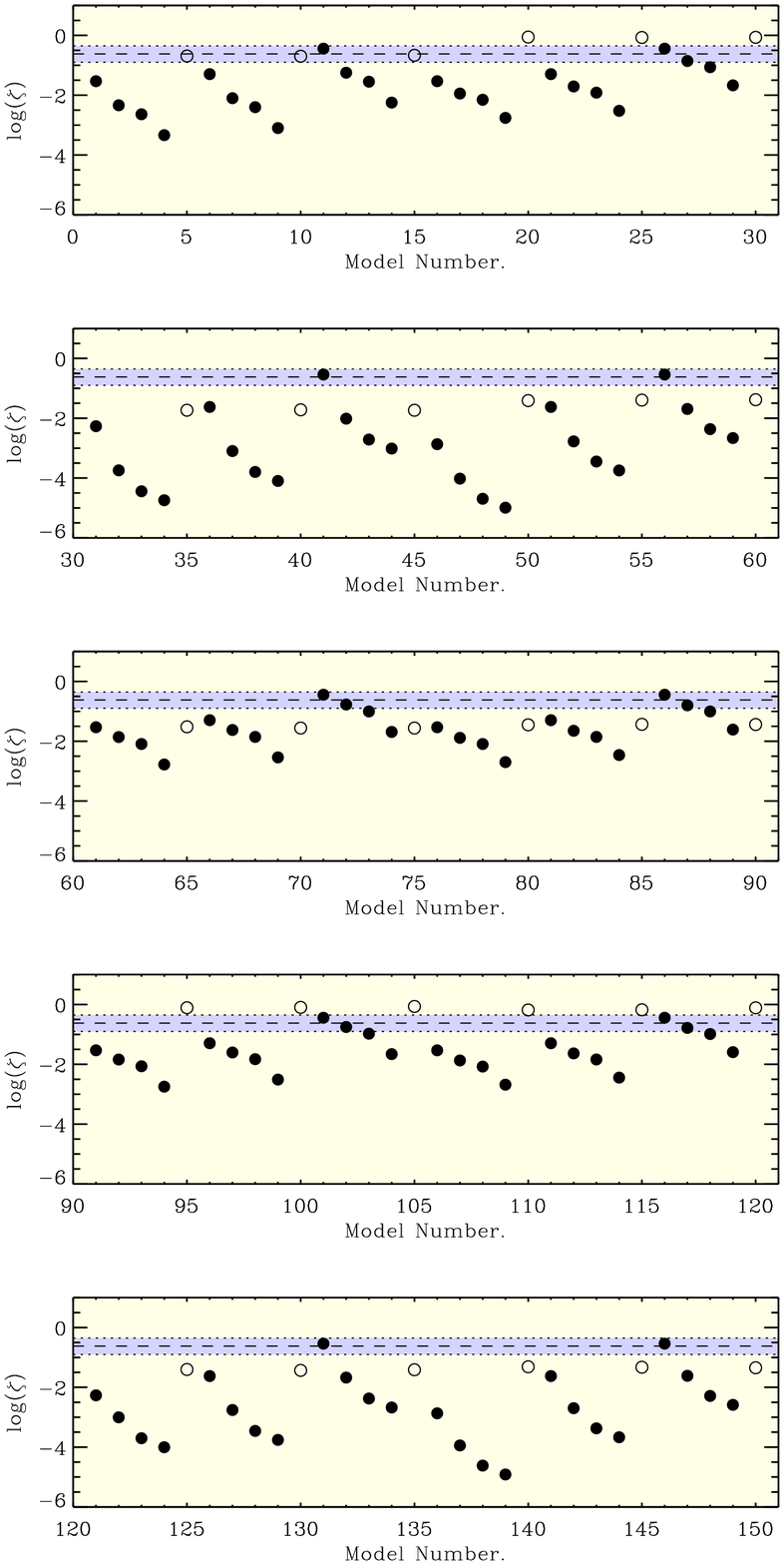}
  }
  \caption{\label{quasar_zeta} Left panels show the dust masses predicted by the 150 parameter configurations considered for the model of
           SDSS J1148+525. Right panels show the dust-to-metals ratio for the same. {See Appendix \ref{grid} for further details and plots of
           the dust mass evolution of each model.}}
  \end{figure*}

\section{Summary and Conclusions}
Stars can only produce a limited amount of dust and the time-scale of survival of dust in {the} ISM may not be long enough.
Thus, there are three fundamental reasons why stellar sources probably cannot explain the dust masses observed at several high-$z$ objects:
\begin{enumerate}
\item Stellar yields must under most circumstances be close to the theoretical maximum for simple dust evolution models to be consistent
      with dust masses derived from observations.
\item A top-heavy IMF means no paricular improvement since the SN-rate increases and therefore the rate of dust destruction increases as well.
      Consequently the dust-to-metals ratio changes in an unfarvourable fashion.
\item Infall of gas (which is more likely than a closed-box scenario) may shorten the dust-destruction time scale and therby lower the effective stellar
      dust production.
\end{enumerate}
The simple analytical models presented in this paper suggest the dust-mass problem in high-$z$ galaxies cannot be solved by only adding the
contribution from AGB stars. The inclusion of AGB stars by made \citet{Valiante09} adds new and important information, but it does not provide a
complete and viable solution to the dust production problem at high redshifts.

\citet{Elvis02} have suggested dust may be created in broad emission line clouds in an outflowing quasar wind. One may also assume dust grains
can form, or at least grow, in the interstellar gas of quasar host galaxies. However, neither hypotheses can properly solve the dust production problem,
because the amount of metals needed as raw-material for dust formation cannot be produced by stars alone in many cases.

{There is essentially} only two ways out of the dilemma described above, is either significant non-stellar dust production (e.g., growth of dust grains in 
the ISM or dust formation associated with quasar outflows) or signifiant systematic errors in the dust and (molecular) gas masses derived from observations. 
{In a few cases it may be sufficient to assume no dust destruction, but this cannot solve the problem in general.}

\section*{Acknowledgments}
The reviewer, Mike Edmunds, is thanked for his valuable comments which helped to improve the paper. Anja C. Andersen and Christa Gall are thanked for
interesting discussions on dust in the early Universe and comments on the contents of this paper. Rosa Valiante is thanked for providing the input data
of her model of SDSS J1148+525. The author acknowledges support from Vetenskapsr\aa det (the Swedish Research Council).
The Dark Cosmology Centre is funded by the Danish National Research Foundation.

\appendix
\section{Models of SDSS J1148+525}
\label{grid}

  \begin{table*}
  \begin{center}
  \caption{\label{parameters} Parameter values for models with a normal IMF and observed gas mass. Masses are given for $z = 6.42$.}
  \begin{tabular}{lllllllllll}

  Model & $\alpha$ & $\epsilon$ & $y_{\rm d}$ & $y_Z$ & $m_{\rm ISM}$ [$M_\odot$] & $\delta_{\rm ISM}$ & $\nu$ & $M_{\rm gas}$ [$M_\odot$] & $M_{\rm tot}$ [$M_\odot$] & Description \\
  \hline
  1     & 0.63     & -          & $8.81\cdot 10^{-4}$ & $3.00\cdot 10^{-2}$ & 0             & 0.00               & -     & $1.6\cdot 10^{10}$ & $8.85\cdot 10^{11}$ & Closed box, no dust destruction.\\
  2     & 0.63     & -          & $8.81\cdot 10^{-4}$ & $3.00\cdot 10^{-2}$ & 100           & 1.00               & 1.59  & $1.6\cdot 10^{10}$ & $8.85\cdot 10^{11}$ & Closed box.\\
  3     & 0.63     & -          & $8.81\cdot 10^{-4}$ & $3.00\cdot 10^{-2}$ & 200           & 2.00               & 3.17  & $1.6\cdot 10^{10}$ & $8.85\cdot 10^{11}$ & Closed box.\\
  4     & 0.63     & -          & $8.81\cdot 10^{-4}$ & $3.00\cdot 10^{-2}$ & 1000          & 10.0               & 15.9  & $1.6\cdot 10^{10}$ & $8.85\cdot 10^{11}$ & Closed box.\\
  5     & 0.63     & 0.35       & $8.81\cdot 10^{-4}$ & $3.00\cdot 10^{-2}$ & 0             & 0.00               & -     & $1.6\cdot 10^{10}$ & $8.85\cdot 10^{11}$ & Closed box, 'secondary dust'.\\[1mm]

  6     & 0.63     & -          & $1.52\cdot 10^{-3}$ & $3.00\cdot 10^{-2}$ & 0             & 0.00               & -     & $1.6\cdot 10^{10}$ & $8.85\cdot 10^{11}$ & Closed box, no dust destruction.\\
  7     & 0.63     & -          & $1.52\cdot 10^{-3}$ & $3.00\cdot 10^{-2}$ & 100           & 1.00               & 1.59  & $1.6\cdot 10^{10}$ & $8.85\cdot 10^{11}$ & Closed box.\\
  8     & 0.63     & -          & $1.52\cdot 10^{-3}$ & $3.00\cdot 10^{-2}$ & 200           & 2.00               & 3.17  & $1.6\cdot 10^{10}$ & $8.85\cdot 10^{11}$ & Closed box.\\
  9     & 0.63     & -          & $1.52\cdot 10^{-3}$ & $3.00\cdot 10^{-2}$ & 1000          & 10.0               & 15.9  & $1.6\cdot 10^{10}$ & $8.85\cdot 10^{11}$ & Closed box.\\
  10    & 0.63     & 0.30       & $1.52\cdot 10^{-3}$ & $3.00\cdot 10^{-2}$ & 0             & 0.00               & -     & $1.6\cdot 10^{10}$ & $8.85\cdot 10^{11}$ & Closed box, 'secondary dust'.\\[1mm]
  
  11    & 0.63     & -          & $1.08\cdot 10^{-2}$ & $3.00\cdot 10^{-2}$ & 0             & 0.00               & -     & $1.6\cdot 10^{10}$ & $8.85\cdot 10^{11}$ & Closed box, no dust destruction.\\
  12    & 0.63     & -          & $1.08\cdot 10^{-2}$ & $3.00\cdot 10^{-2}$ & 100           & 1.00               & 1.59  & $1.6\cdot 10^{10}$ & $8.85\cdot 10^{11}$ & Closed box.\\                     
  13    & 0.63     & -          & $1.08\cdot 10^{-2}$ & $3.00\cdot 10^{-2}$ & 200           & 2.00               & 3.17  & $1.6\cdot 10^{10}$ & $8.85\cdot 10^{11}$ & Closed box.\\                     
  14    & 0.63     & -          & $1.08\cdot 10^{-2}$ & $3.00\cdot 10^{-2}$ & 1000          & 10.0               & 15.9  & $1.6\cdot 10^{10}$ & $8.85\cdot 10^{11}$ & Closed box.\\
  15    & 0.63     & 0.30       & $1.08\cdot 10^{-2}$ & $3.00\cdot 10^{-2}$ & 100           & 1.00               & 1.59  & $1.6\cdot 10^{10}$ & $8.85\cdot 10^{11}$ & Closed box, 'secondary dust'.\\[1mm]
 
  16    & 0.63     & -          & $8.81\cdot 10^{-4}$ & $3.00\cdot 10^{-2}$ & 0             & 0.00               & -     & $1.6\cdot 10^{10}$ & $8.85\cdot 10^{11}$ & Infall, no dust destruction.\\
  17    & 0.63     & -          & $8.81\cdot 10^{-4}$ & $3.00\cdot 10^{-2}$ & 100           & 1.00               & 1.59  & $1.6\cdot 10^{10}$ & $8.85\cdot 10^{11}$ & Infall\\                     
  18    & 0.63     & -          & $8.81\cdot 10^{-4}$ & $3.00\cdot 10^{-2}$ & 200           & 2.00               & 3.17  & $1.6\cdot 10^{10}$ & $8.85\cdot 10^{11}$ & Infall\\
  19    & 0.63     & -          & $8.81\cdot 10^{-4}$ & $3.00\cdot 10^{-2}$ & 1000          & 10.0               & 15.9  & $1.6\cdot 10^{10}$ & $8.85\cdot 10^{11}$ & Infall\\
  20    & 0.63     & 0.85       & $8.81\cdot 10^{-4}$ & $3.00\cdot 10^{-2}$ & 0             & 0.00               & -     & $1.6\cdot 10^{10}$ & $8.85\cdot 10^{11}$ & Infall, 'secondary dust'.\\[1mm]
  
  21    & 0.63     & -          & $1.52\cdot 10^{-3}$ & $3.00\cdot 10^{-2}$ & 0             & 0.00               & -     & $1.6\cdot 10^{10}$ & $8.85\cdot 10^{11}$ & Infall, no dust destruction.\\
  22    & 0.63     & -          & $1.52\cdot 10^{-3}$ & $3.00\cdot 10^{-2}$ & 100           & 1.00               & 1.59  & $1.6\cdot 10^{10}$ & $8.85\cdot 10^{11}$ & Infall\\                      
  23    & 0.63     & -          & $1.52\cdot 10^{-3}$ & $3.00\cdot 10^{-2}$ & 200           & 2.00               & 3.17  & $1.6\cdot 10^{10}$ & $8.85\cdot 10^{11}$ & Infall\\                      
  24    & 0.63     & -          & $1.52\cdot 10^{-3}$ & $3.00\cdot 10^{-2}$ & 1000          & 10.0               & 15.9  & $1.6\cdot 10^{10}$ & $8.85\cdot 10^{11}$ & Infall\\                      
  25    & 0.63     & 0.80       & $1.52\cdot 10^{-3}$ & $3.00\cdot 10^{-2}$ & 0             & 0.00               & -     & $1.6\cdot 10^{10}$ & $8.85\cdot 10^{11}$ & Infall, 'secondary dust'.\\[1mm]
  
  26    & 0.63     & -          & $1.08\cdot 10^{-2}$ & $3.00\cdot 10^{-2}$ & 0             & 0.00               & -     & $1.6\cdot 10^{10}$ & $8.85\cdot 10^{11}$ & Infall, no dust destruction.\\
  27    & 0.63     & -          & $1.08\cdot 10^{-2}$ & $3.00\cdot 10^{-2}$ & 100           & 1.00               & 1.59  & $1.6\cdot 10^{10}$ & $8.85\cdot 10^{11}$ & Infall\\
  28    & 0.63     & -          & $1.08\cdot 10^{-2}$ & $3.00\cdot 10^{-2}$ & 200           & 2.00               & 3.17  & $1.6\cdot 10^{10}$ & $8.85\cdot 10^{11}$ & Infall\\
  29    & 0.63     & -          & $1.08\cdot 10^{-2}$ & $3.00\cdot 10^{-2}$ & 1000          & 10.0               & 15.9  & $1.6\cdot 10^{10}$ & $8.85\cdot 10^{11}$ & Infall\\
  30    & 0.63     & 0.50       & $1.08\cdot 10^{-2}$ & $3.00\cdot 10^{-2}$ & 0             & 0.00               & -     & $1.6\cdot 10^{10}$ & $8.85\cdot 10^{11}$ & Infall, 'secondary dust'.\\
  \hline
  \end{tabular}
  \end{center}
  \end{table*}

  \begin{figure*}

  \resizebox{15.9cm}{!}{
  \includegraphics{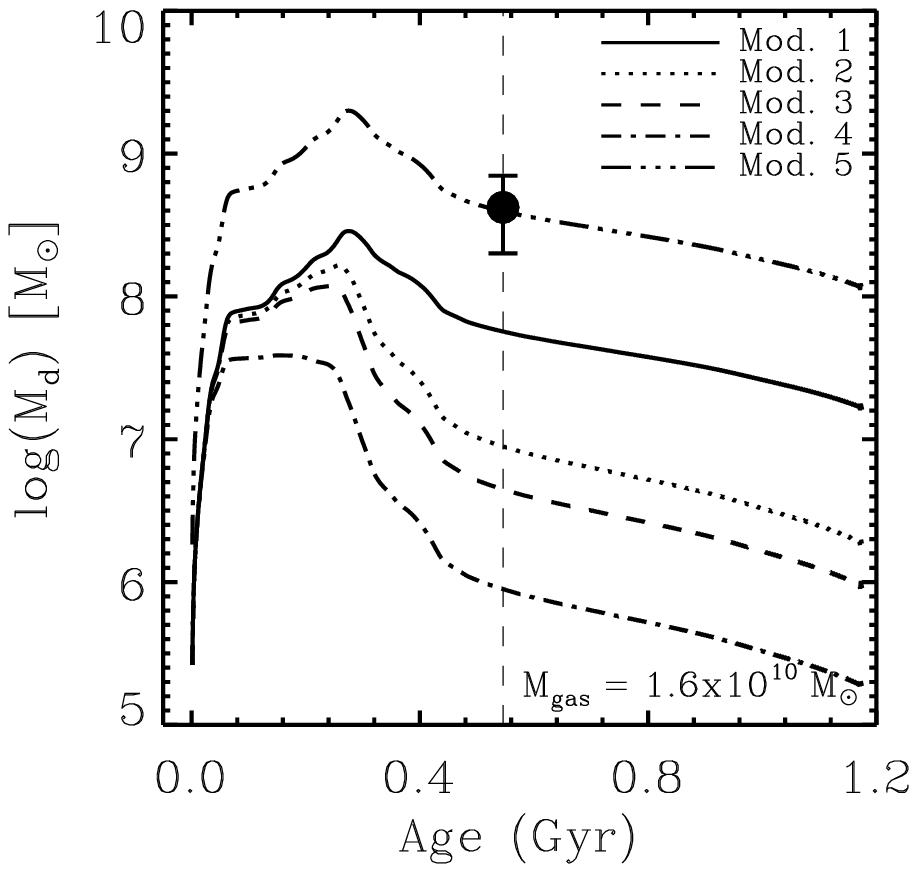}
  \includegraphics{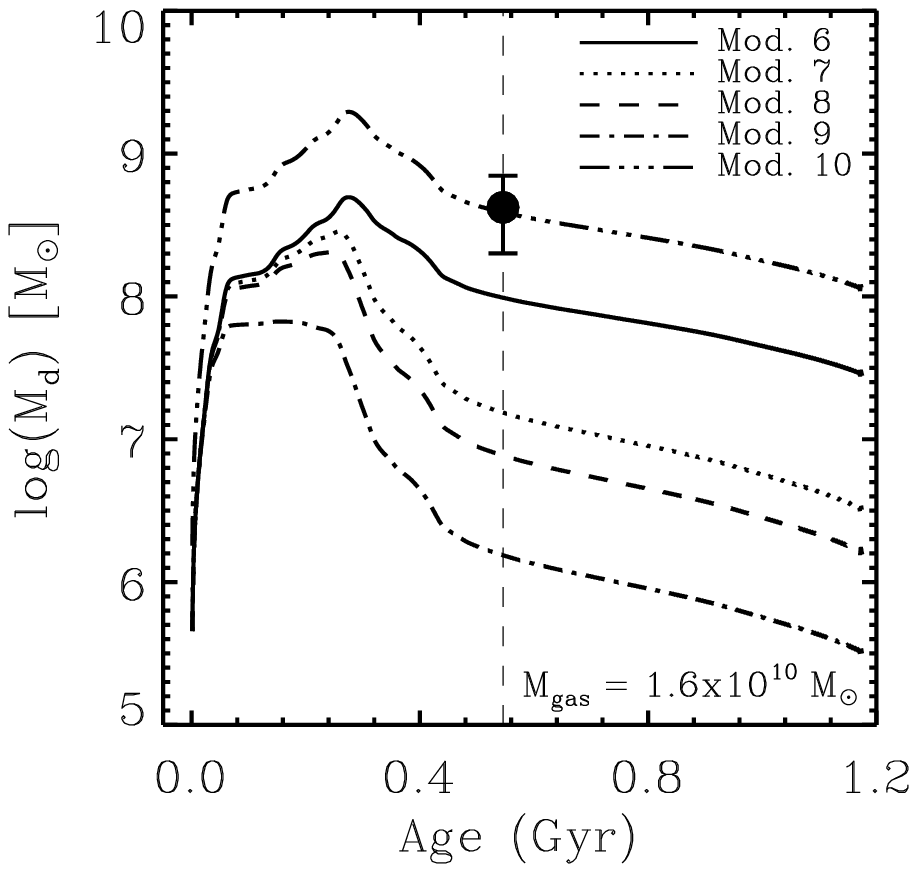}
  \includegraphics{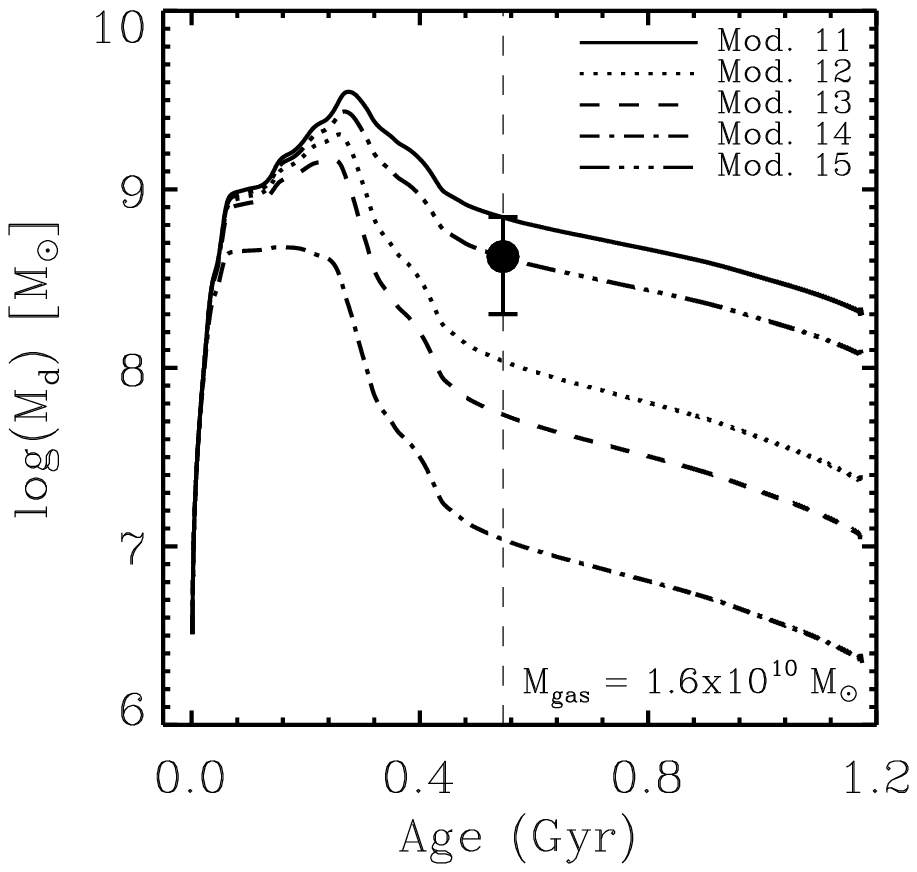}
  }
  \resizebox{15.9cm}{!}{
  \includegraphics{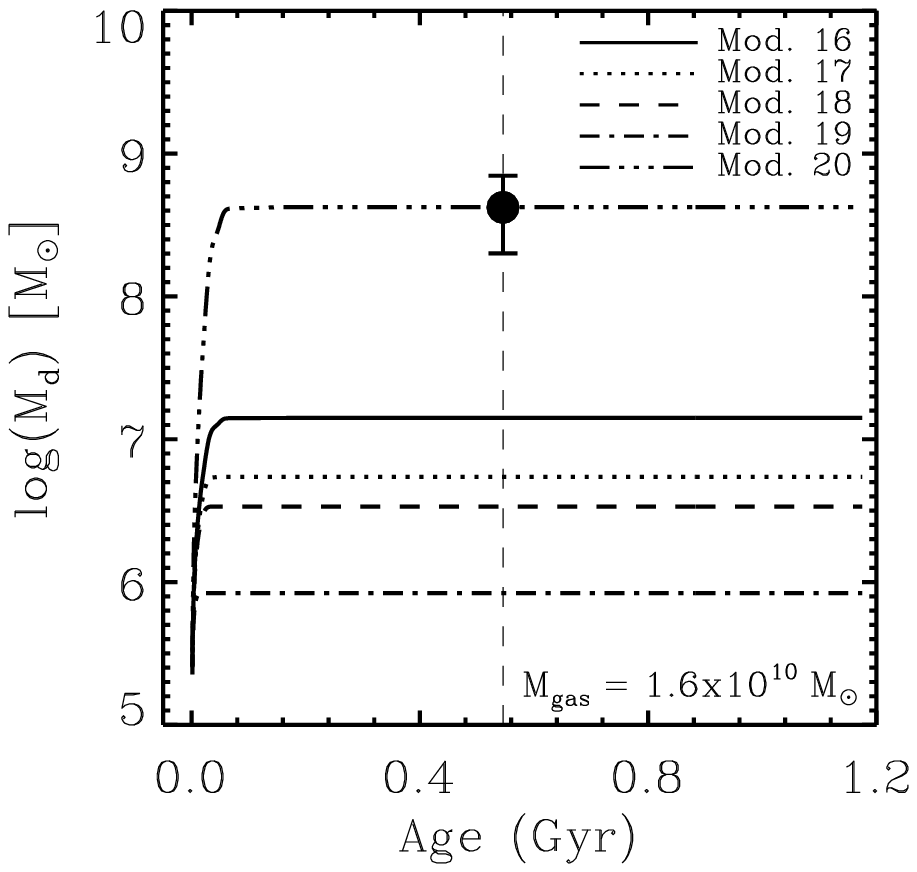}
  \includegraphics{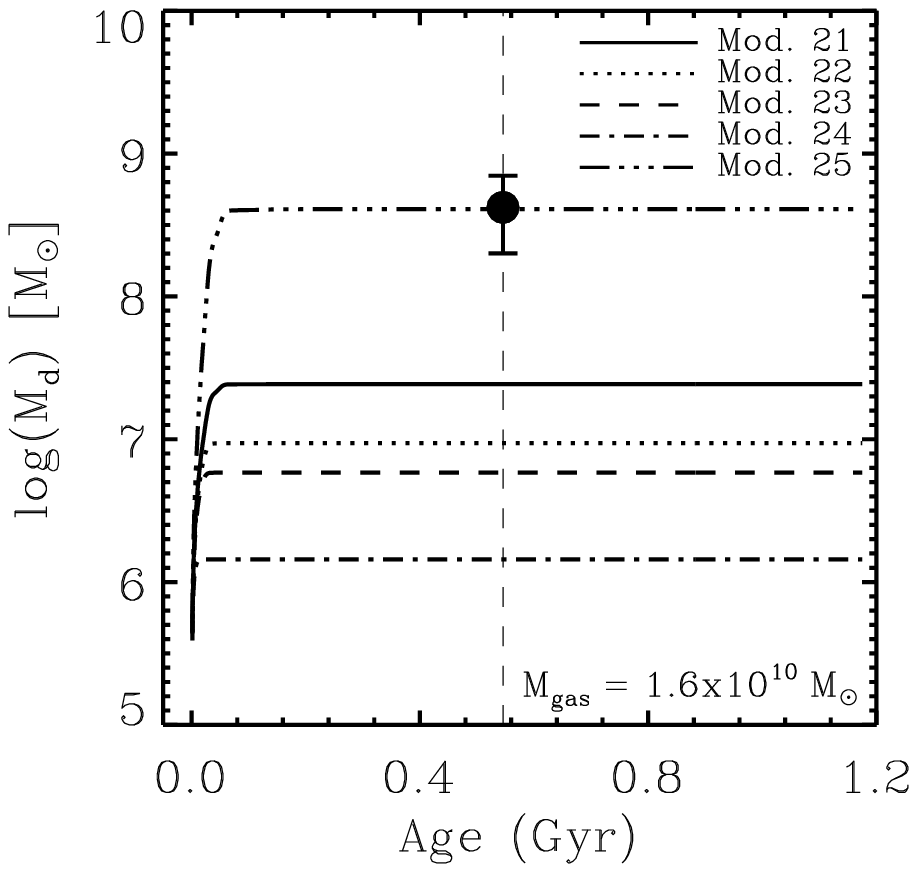}
  \includegraphics{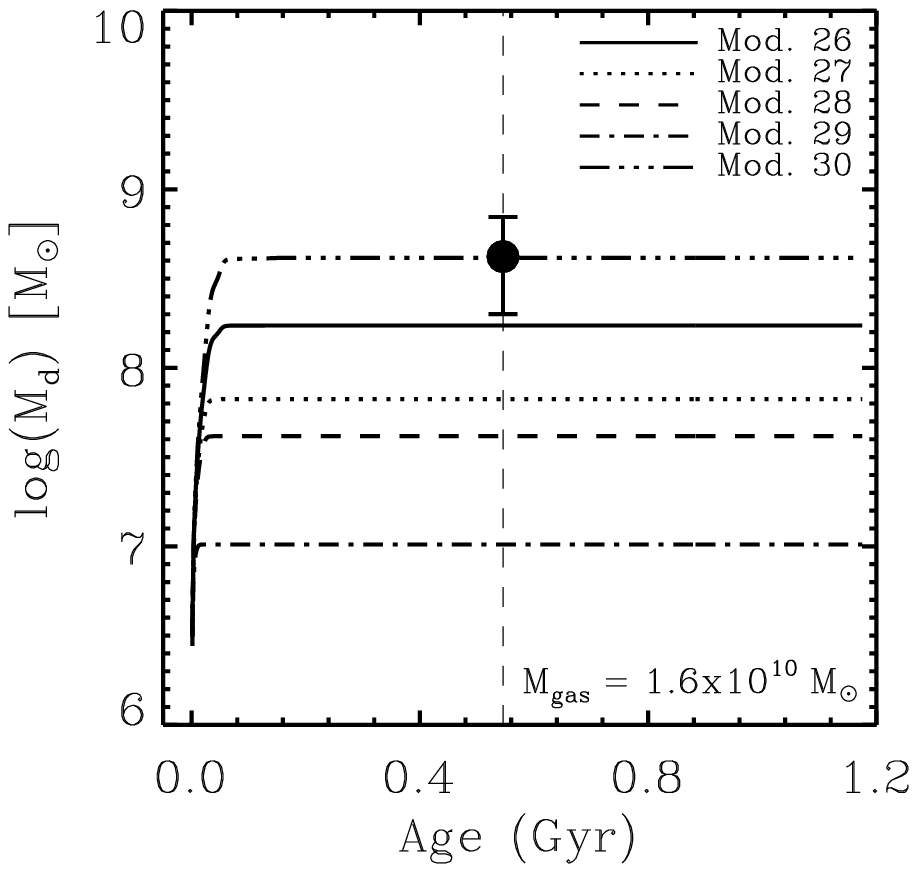}
  }
  \caption{Models using the star-formation history resulting from the simulation of Li et al. (2007). The
           filled circle with error bars show the estimated dust mass of SDSS J1148+525.
          \label{quasar}}
  \end{figure*}

    \begin{table*}
  \begin{center}
  \caption{\label{parameters_th} Parameter values for models with a top-heavy IMF and observed gas mass. Masses are given for $z = 6.42$.}
  \begin{tabular}{lllllllllll}

  Model & $\alpha$ & $\epsilon$ & $y_{\rm d}$ & $y_Z$ & $m_{\rm ISM}$ [$M_\odot$] & $\delta_{\rm ISM}$ & $\nu$ & $M_{\rm gas}$ [$M_\odot$] & $M_{\rm tot}$ [$M_\odot$] & Description \\
  \hline
  31    & 0.10     & -          & $3.51\cdot 10^{-3}$ & $6.50\cdot 10^{-1}$ & 0             & 0.00               & -     & $1.6\cdot 10^{10}$ & $1.54\cdot 10^{11}$ & Closed box, no dust destruction.\\
  32    & 0.10     & -          & $3.51\cdot 10^{-3}$ & $6.50\cdot 10^{-1}$ & 20            & 1.32               & 1.59  & $1.6\cdot 10^{10}$ & $1.54\cdot 10^{11}$ & Closed box.\\
  33    & 0.10     & -          & $3.51\cdot 10^{-3}$ & $6.50\cdot 10^{-1}$ & 100           & 6.60               & 3.17  & $1.6\cdot 10^{10}$ & $1.54\cdot 10^{11}$ & Closed box.\\
  34    & 0.10     & -          & $3.51\cdot 10^{-3}$ & $6.50\cdot 10^{-1}$ & 200           & 13.2               & 15.9  & $1.6\cdot 10^{10}$ & $1.54\cdot 10^{11}$ & Closed box.\\
  35    & 0.10     & 0.25       & $3.51\cdot 10^{-3}$ & $6.50\cdot 10^{-1}$ & 20            & 1.32               & -     & $1.6\cdot 10^{10}$ & $1.54\cdot 10^{11}$ & Closed box, 'secondary dust'.\\[1mm]
                                       
  36    & 0.10     & -          & $1.55\cdot 10^{-2}$ & $6.50\cdot 10^{-1}$ & 0             & 0.00               & -     & $1.6\cdot 10^{10}$ & $1.54\cdot 10^{11}$ & Closed box, no dust destruction.\\
  37    & 0.10     & -          & $1.55\cdot 10^{-2}$ & $6.50\cdot 10^{-1}$ & 20            & 1.32               & 1.59  & $1.6\cdot 10^{10}$ & $1.54\cdot 10^{11}$ & Closed box.\\
  38    & 0.10     & -          & $1.55\cdot 10^{-2}$ & $6.50\cdot 10^{-1}$ & 100           & 6.60               & 3.17  & $1.6\cdot 10^{10}$ & $1.54\cdot 10^{11}$ & Closed box.\\
  39    & 0.10     & -          & $1.55\cdot 10^{-2}$ & $6.50\cdot 10^{-1}$ & 200           & 13.2               & 15.9  & $1.6\cdot 10^{10}$ & $1.54\cdot 10^{11}$ & Closed box.\\
  40    & 0.10     & 0.25       & $1.55\cdot 10^{-2}$ & $6.50\cdot 10^{-1}$ & 20            & 1.32               & -     & $1.6\cdot 10^{10}$ & $1.54\cdot 10^{11}$ & Closed box, 'secondary dust'.\\[1mm]
                                       
  41    & 0.10     & -          & $1.88\cdot 10^{-1}$ & $6.50\cdot 10^{-1}$ & 0             & 0.00               & -     & $1.6\cdot 10^{10}$ & $1.54\cdot 10^{11}$ & Closed box, no dust destruction.\\
  42    & 0.10     & -          & $1.88\cdot 10^{-1}$ & $6.50\cdot 10^{-1}$ & 20            & 1.32               & 1.59  & $1.6\cdot 10^{10}$ & $1.54\cdot 10^{11}$ & Closed box.\\
  43    & 0.10     & -          & $1.88\cdot 10^{-1}$ & $6.50\cdot 10^{-1}$ & 100           & 6.60               & 3.17  & $1.6\cdot 10^{10}$ & $1.54\cdot 10^{11}$ & Closed box.\\
  44    & 0.10     & -          & $1.88\cdot 10^{-1}$ & $6.50\cdot 10^{-1}$ & 200           & 13.2               & 15.9  & $1.6\cdot 10^{10}$ & $1.54\cdot 10^{11}$ & Closed box.\\
  45    & 0.10     & 0.12       & $1.88\cdot 10^{-1}$ & $6.50\cdot 10^{-1}$ & 20            & 1.32               & 1.59  & $1.6\cdot 10^{10}$ & $1.54\cdot 10^{11}$ & Closed box, 'secondary dust'.\\[1mm]

  46    & 0.10     & -          & $3.51\cdot 10^{-3}$ & $6.50\cdot 10^{-1}$ & 0             & 0.00               & -     & $1.6\cdot 10^{10}$ & $1.54\cdot 10^{11}$ & Infall, no dust destruction.\\
  47    & 0.10     & -          & $3.51\cdot 10^{-3}$ & $6.50\cdot 10^{-1}$ & 20            & 1.32               & 1.59  & $1.6\cdot 10^{10}$ & $1.54\cdot 10^{11}$ & Infall\\
  48    & 0.10     & -          & $3.51\cdot 10^{-3}$ & $6.50\cdot 10^{-1}$ & 100           & 6.60               & 3.17  & $1.6\cdot 10^{10}$ & $1.54\cdot 10^{11}$ & Infall\\
  49    & 0.10     & -          & $3.51\cdot 10^{-3}$ & $6.50\cdot 10^{-1}$ & 200           & 13.2               & 15.9  & $1.6\cdot 10^{10}$ & $1.54\cdot 10^{11}$ & Infall\\
  50    & 0.10     & 0.55       & $3.51\cdot 10^{-3}$ & $6.50\cdot 10^{-1}$ & 20            & 1.32               & -     & $1.6\cdot 10^{10}$ & $1.54\cdot 10^{11}$ & Infall, 'secondary dust'.\\[1mm]
                                                    
  51    & 0.10     & -          & $1.55\cdot 10^{-2}$ & $6.50\cdot 10^{-1}$ & 0             & 0.00               & -     & $1.6\cdot 10^{10}$ & $1.54\cdot 10^{11}$ & Infall, no dust destruction.\\
  52    & 0.10     & -          & $1.55\cdot 10^{-2}$ & $6.50\cdot 10^{-1}$ & 20            & 1.32               & 1.59  & $1.6\cdot 10^{10}$ & $1.54\cdot 10^{11}$ & Infall\\
  53    & 0.10     & -          & $1.55\cdot 10^{-2}$ & $6.50\cdot 10^{-1}$ & 100           & 6.60               & 3.17  & $1.6\cdot 10^{10}$ & $1.54\cdot 10^{11}$ & Infall\\
  54    & 0.10     & -          & $1.55\cdot 10^{-2}$ & $6.50\cdot 10^{-1}$ & 200           & 13.2               & 15.9  & $1.6\cdot 10^{10}$ & $1.54\cdot 10^{11}$ & Infall\\
  55    & 0.10     & 0.55       & $1.55\cdot 10^{-2}$ & $6.50\cdot 10^{-1}$ & 20            & 1.32               & -     & $1.6\cdot 10^{10}$ & $1.54\cdot 10^{11}$ & Infall, 'secondary dust'.\\[1mm]

  56    & 0.10     & -          & $1.88\cdot 10^{-1}$ & $6.50\cdot 10^{-1}$ & 0             & 0.00               & -     & $1.6\cdot 10^{10}$ & $1.54\cdot 10^{11}$ & Infall, no dust destruction.\\
  57    & 0.10     & -          & $1.88\cdot 10^{-1}$ & $6.50\cdot 10^{-1}$ & 20            & 1.32               & 1.59  & $1.6\cdot 10^{10}$ & $1.54\cdot 10^{11}$ & Infall\\
  58    & 0.10     & -          & $1.88\cdot 10^{-1}$ & $6.50\cdot 10^{-1}$ & 100           & 6.60               & 3.17  & $1.6\cdot 10^{10}$ & $1.54\cdot 10^{11}$ & Infall\\
  59    & 0.10     & -          & $1.88\cdot 10^{-1}$ & $6.50\cdot 10^{-1}$ & 200           & 13.2               & 15.9  & $1.6\cdot 10^{10}$ & $1.54\cdot 10^{11}$ & Infall\\
  60    & 0.10     & 0.30       & $1.88\cdot 10^{-1}$ & $6.50\cdot 10^{-1}$ & 20            & 1.32               & -     & $1.6\cdot 10^{10}$ & $1.54\cdot 10^{11}$ & Infall, 'secondary dust'.\\
  \hline
  \end{tabular}
  \end{center}
  \end{table*}

  \begin{figure*}

  \resizebox{15.9cm}{!}{
  \includegraphics{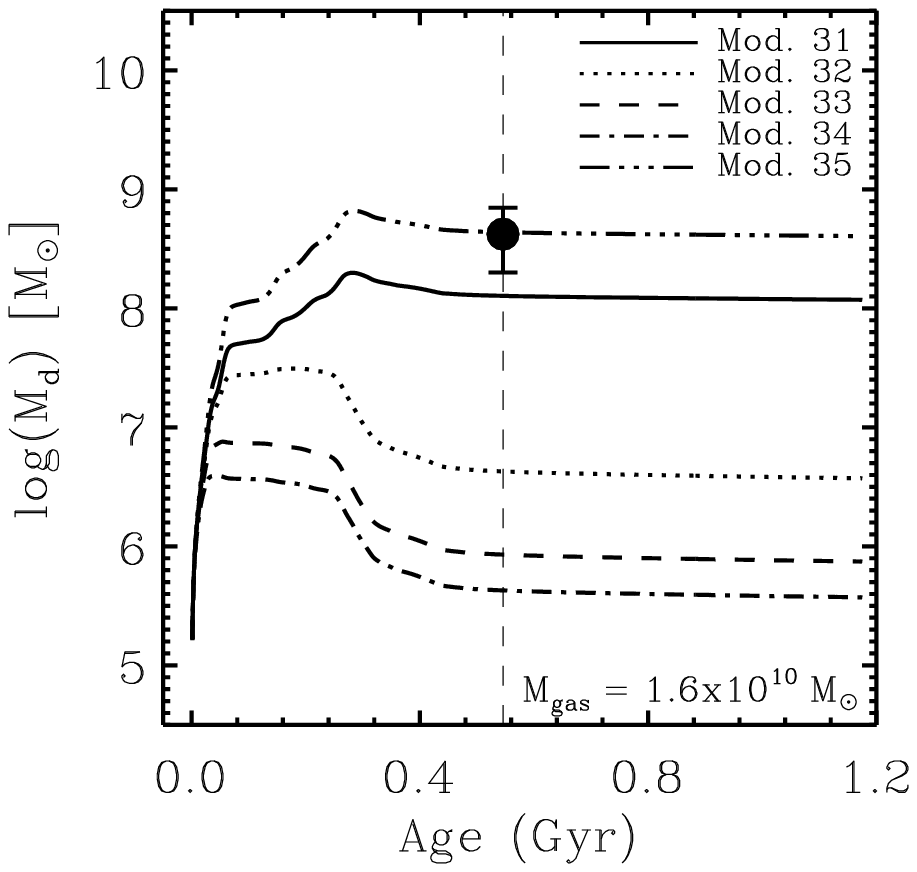}
  \includegraphics{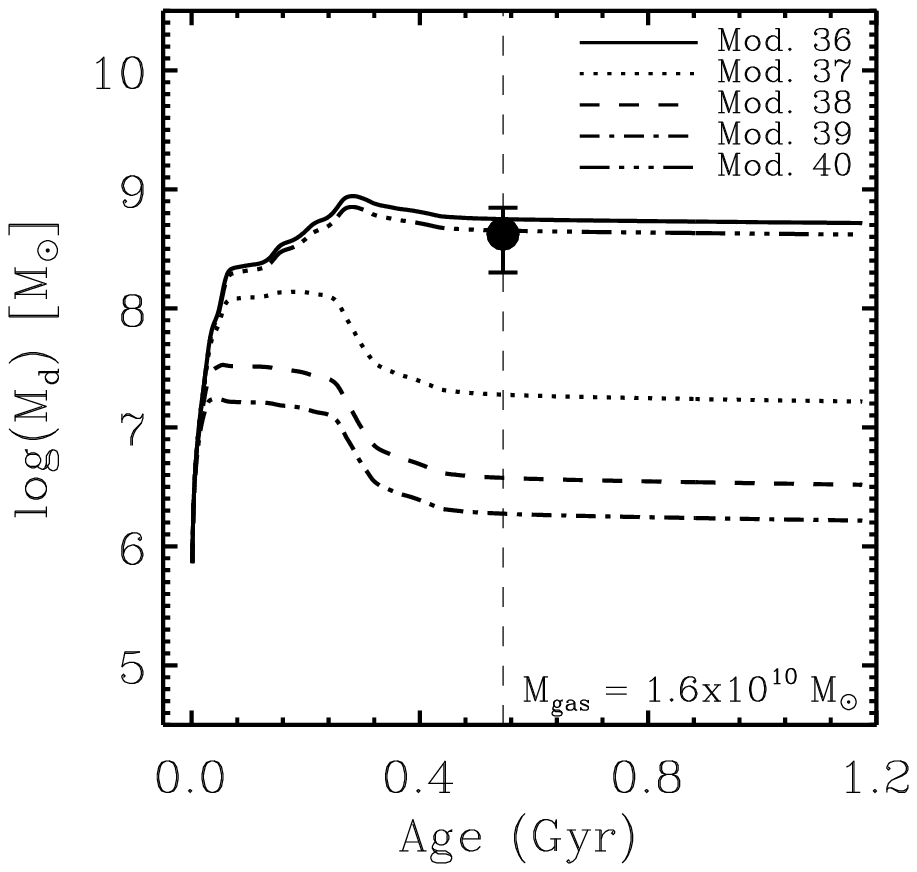}
  \includegraphics{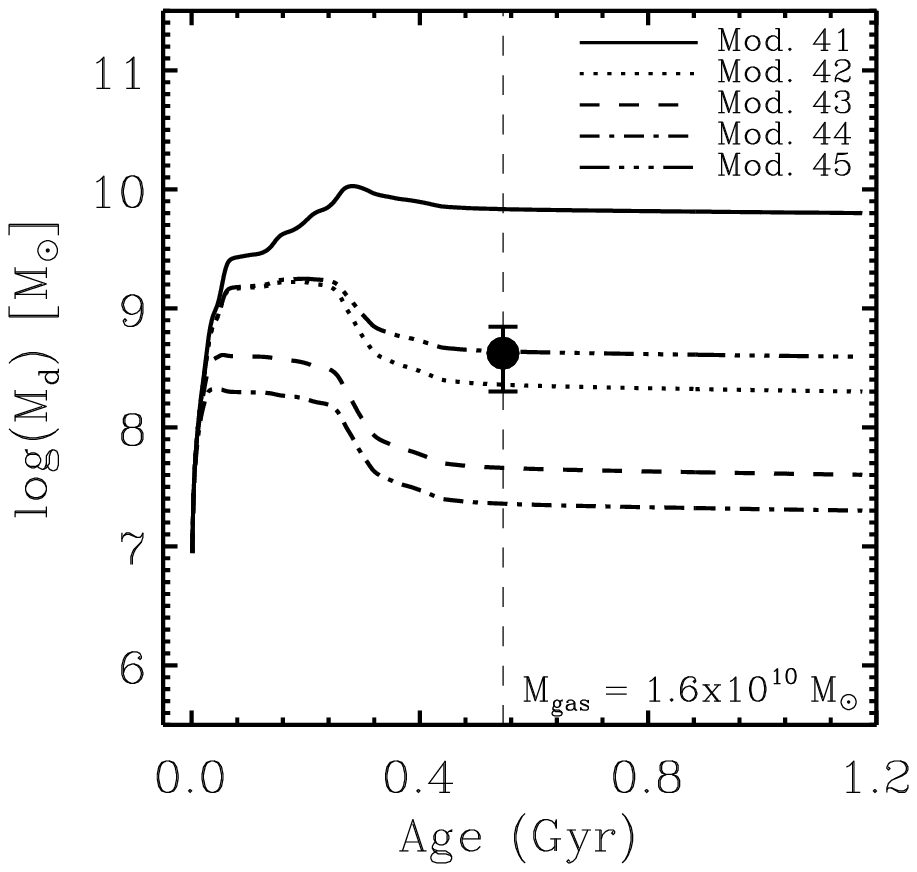}
  }
  \resizebox{15.9cm}{!}{
  \includegraphics{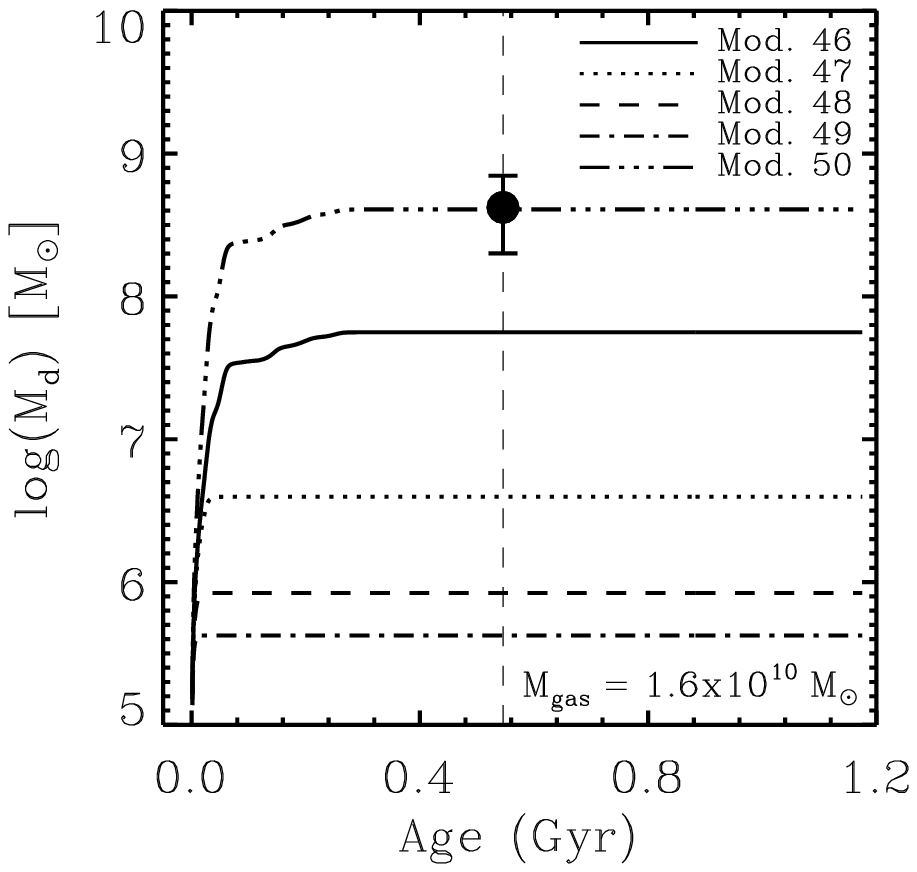}
  \includegraphics{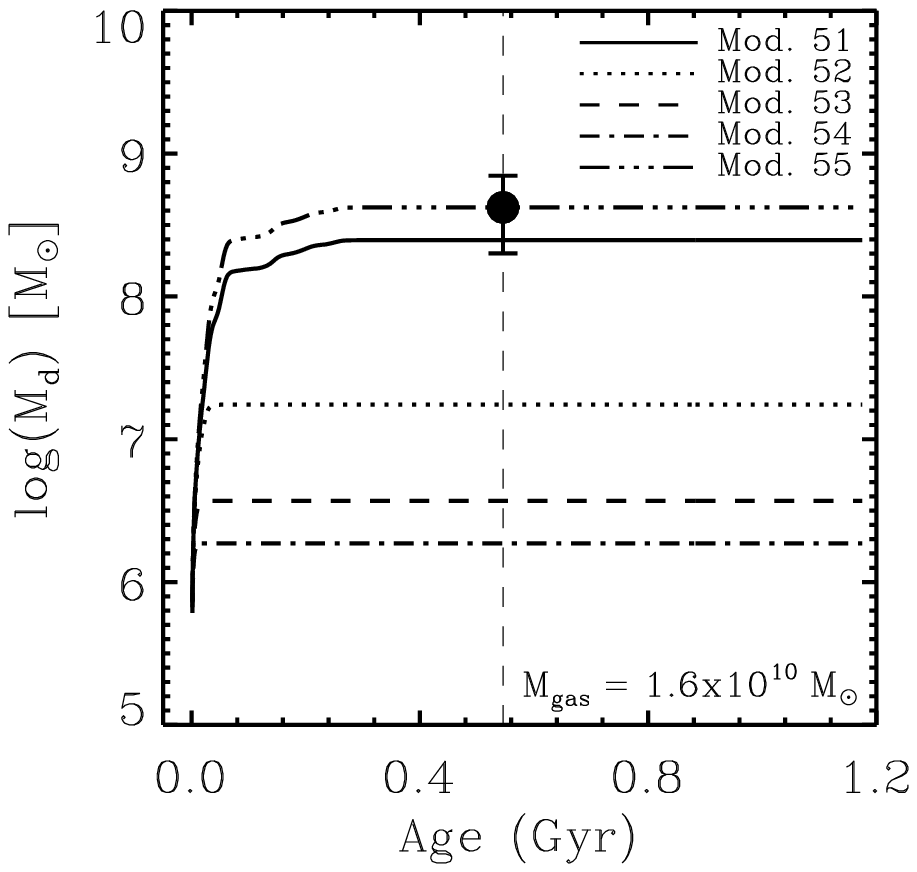}
  \includegraphics{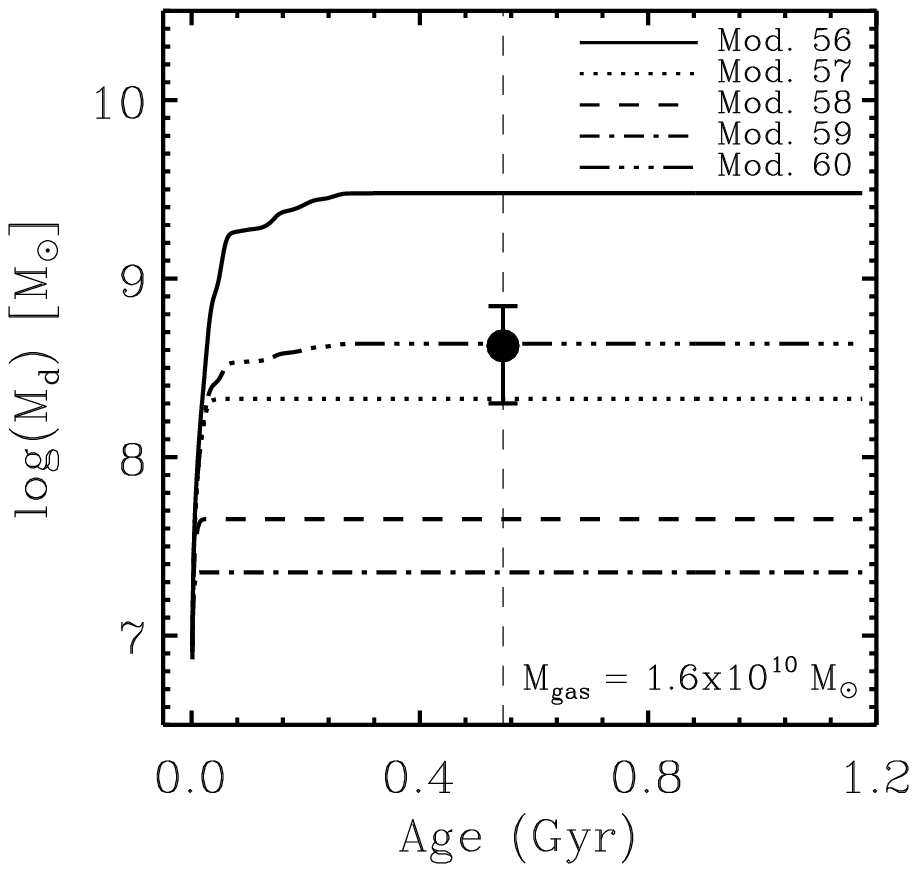}
  }
  \caption{Same as Fig. \ref{quasar}, but for the models with a top-heavy IMF.
          \label{quasar_th}}
  \end{figure*}

    \begin{table*}
  \begin{center}
  \caption{\label{parameters_g} Parameter values for models with a normal IMF and a total mass of $1.3\cdot 10^{12}M_\odot$ \citep{Li07, Valiante09}.
                              Masses are given for $z = 6.42$.}
  \begin{tabular}{lllllllllll}

  Model & $\alpha$ & $\epsilon$ & $y_{\rm d}$ & $y_Z$ & $m_{\rm ISM}$ [$M_\odot$] & $\delta_{\rm ISM}$ & $\nu$ & $M_{\rm gas}$ [$M_\odot$] & $M_{\rm tot}$ [$M_\odot$] & Description \\
  \hline
  61    & 0.63     & -          & $8.81\cdot 10^{-4}$ & $3.00\cdot 10^{-2}$ & 0             & 0.00               & -     & $4.3\cdot 10^{11}$ & $1.30\cdot 10^{12}$ & Closed box, no dust destruction.\\
  62    & 0.63     & -          & $8.81\cdot 10^{-4}$ & $3.00\cdot 10^{-2}$ & 100           & 1.00               & 1.59  & $4.3\cdot 10^{11}$ & $1.30\cdot 10^{12}$ & Closed box.\\
  63    & 0.63     & -          & $8.81\cdot 10^{-4}$ & $3.00\cdot 10^{-2}$ & 200           & 2.00               & 3.17  & $4.3\cdot 10^{11}$ & $1.30\cdot 10^{12}$ & Closed box.\\
  64    & 0.63     & -          & $8.81\cdot 10^{-4}$ & $3.00\cdot 10^{-2}$ & 1000          & 10.0               & 15.9  & $4.3\cdot 10^{11}$ & $1.30\cdot 10^{12}$ & Closed box.\\
  65    & 0.63     & 0.05       & $8.81\cdot 10^{-4}$ & $3.00\cdot 10^{-2}$ & 100           & 1.00               & 1.59  & $4.3\cdot 10^{11}$ & $1.30\cdot 10^{12}$ & Closed box, 'secondary dust'.\\[1mm]

  66    & 0.63     & -          & $1.52\cdot 10^{-3}$ & $3.00\cdot 10^{-2}$ & 0             & 0.00               & -     & $4.3\cdot 10^{11}$ & $1.30\cdot 10^{12}$ & Closed box, no dust destruction.\\
  67    & 0.63     & -          & $1.52\cdot 10^{-3}$ & $3.00\cdot 10^{-2}$ & 100           & 1.00               & 1.59  & $4.3\cdot 10^{11}$ & $1.30\cdot 10^{12}$ & Closed box.\\
  68    & 0.63     & -          & $1.52\cdot 10^{-3}$ & $3.00\cdot 10^{-2}$ & 200           & 2.00               & 3.17  & $4.3\cdot 10^{11}$ & $1.30\cdot 10^{12}$ & Closed box.\\
  69    & 0.63     & -          & $1.52\cdot 10^{-3}$ & $3.00\cdot 10^{-2}$ & 1000          & 10.0               & 15.9  & $4.3\cdot 10^{11}$ & $1.30\cdot 10^{12}$ & Closed box.\\
  70    & 0.63     & 0.06       & $1.52\cdot 10^{-3}$ & $3.00\cdot 10^{-2}$ & 200           & 2.00               & 3.17  & $4.3\cdot 10^{11}$ & $1.30\cdot 10^{12}$ & Closed box, 'secondary dust'.\\[1mm]

  71    & 0.63     & -          & $1.08\cdot 10^{-2}$ & $3.00\cdot 10^{-2}$ & 0             & 0.00               & -     & $4.3\cdot 10^{11}$ & $1.30\cdot 10^{12}$ & Closed box, no dust destruction.\\
  72    & 0.63     & -          & $1.08\cdot 10^{-2}$ & $3.00\cdot 10^{-2}$ & 100           & 1.00               & 1.59  & $4.3\cdot 10^{11}$ & $1.30\cdot 10^{12}$ & Closed box.\\
  73    & 0.63     & -          & $1.08\cdot 10^{-2}$ & $3.00\cdot 10^{-2}$ & 200           & 2.00               & 3.17  & $4.3\cdot 10^{11}$ & $1.30\cdot 10^{12}$ & Closed box.\\
  74    & 0.63     & -          & $1.08\cdot 10^{-2}$ & $3.00\cdot 10^{-2}$ & 1000          & 10.0               & 15.9  & $4.3\cdot 10^{11}$ & $1.30\cdot 10^{12}$ & Closed box.\\
  75    & 0.63     & 0.12       & $1.08\cdot 10^{-2}$ & $3.00\cdot 10^{-2}$ & 1000          & 10.0               & 15.9  & $4.3\cdot 10^{11}$ & $1.30\cdot 10^{12}$ & Closed box, 'secondary dust'.\\[1mm]

  76    & 0.63     & -          & $8.81\cdot 10^{-4}$ & $3.00\cdot 10^{-2}$ & 0             & 0.00               & -     & $4.3\cdot 10^{11}$ & $1.30\cdot 10^{12}$ & Infall, no dust destruction.\\
  77    & 0.63     & -          & $8.81\cdot 10^{-4}$ & $3.00\cdot 10^{-2}$ & 100           & 1.00               & 1.59  & $4.3\cdot 10^{11}$ & $1.30\cdot 10^{12}$ & Infall\\
  78    & 0.63     & -          & $8.81\cdot 10^{-4}$ & $3.00\cdot 10^{-2}$ & 200           & 2.00               & 3.17  & $4.3\cdot 10^{11}$ & $1.30\cdot 10^{12}$ & Infall\\
  79    & 0.63     & -          & $8.81\cdot 10^{-4}$ & $3.00\cdot 10^{-2}$ & 1000          & 10.0               & 15.9  & $4.3\cdot 10^{11}$ & $1.30\cdot 10^{12}$ & Infall\\
  80    & 0.63     & 0.12       & $8.81\cdot 10^{-4}$ & $3.00\cdot 10^{-2}$ & 200           & 2.00               & 3.17  & $4.3\cdot 10^{11}$ & $1.30\cdot 10^{12}$ & Infall, 'secondary dust'.\\[1mm]

  81    & 0.63     & -          & $1.52\cdot 10^{-3}$ & $3.00\cdot 10^{-2}$ & 0             & 0.00               & -     & $4.3\cdot 10^{11}$ & $1.30\cdot 10^{12}$ & Infall, no dust destruction.\\
  82    & 0.63     & -          & $1.52\cdot 10^{-3}$ & $3.00\cdot 10^{-2}$ & 100           & 1.00               & 1.59  & $4.3\cdot 10^{11}$ & $1.30\cdot 10^{12}$ & Infall\\
  83    & 0.63     & -          & $1.52\cdot 10^{-3}$ & $3.00\cdot 10^{-2}$ & 200           & 2.00               & 3.17  & $4.3\cdot 10^{11}$ & $1.30\cdot 10^{12}$ & Infall\\
  84    & 0.63     & -          & $1.52\cdot 10^{-3}$ & $3.00\cdot 10^{-2}$ & 1000          & 10.0               & 15.9  & $4.3\cdot 10^{11}$ & $1.30\cdot 10^{12}$ & Infall\\
  85    & 0.63     & 0.10       & $1.52\cdot 10^{-3}$ & $3.00\cdot 10^{-2}$ & 200           & 2.00               & 3.17  & $4.3\cdot 10^{11}$ & $1.30\cdot 10^{12}$ & Infall, 'secondary dust'.\\[1mm]

  86    & 0.63     & -          & $1.08\cdot 10^{-2}$ & $3.00\cdot 10^{-2}$ & 0             & 0.00               & -     & $4.3\cdot 10^{11}$ & $1.30\cdot 10^{12}$ & Infall, no dust destruction.\\
  87    & 0.63     & -          & $1.08\cdot 10^{-2}$ & $3.00\cdot 10^{-2}$ & 100           & 1.00               & 1.59  & $4.3\cdot 10^{11}$ & $1.30\cdot 10^{12}$ & Infall\\
  88    & 0.63     & -          & $1.08\cdot 10^{-2}$ & $3.00\cdot 10^{-2}$ & 200           & 2.00               & 3.17  & $4.3\cdot 10^{11}$ & $1.30\cdot 10^{12}$ & Infall\\
  89    & 0.63     & -          & $1.08\cdot 10^{-2}$ & $3.00\cdot 10^{-2}$ & 1000          & 10.0               & 15.9  & $4.3\cdot 10^{11}$ & $1.30\cdot 10^{12}$ & Infall\\
  90    & 0.63     & 0.20       & $1.08\cdot 10^{-2}$ & $3.00\cdot 10^{-2}$ & 1000          & 10.0               & 15.9  & $4.3\cdot 10^{11}$ & $1.30\cdot 10^{12}$ & Infall, 'secondary dust'.\\
  \hline
  \end{tabular}
  \end{center}
  \end{table*}

  \begin{figure*}

  \resizebox{15.9cm}{!}{
  \includegraphics{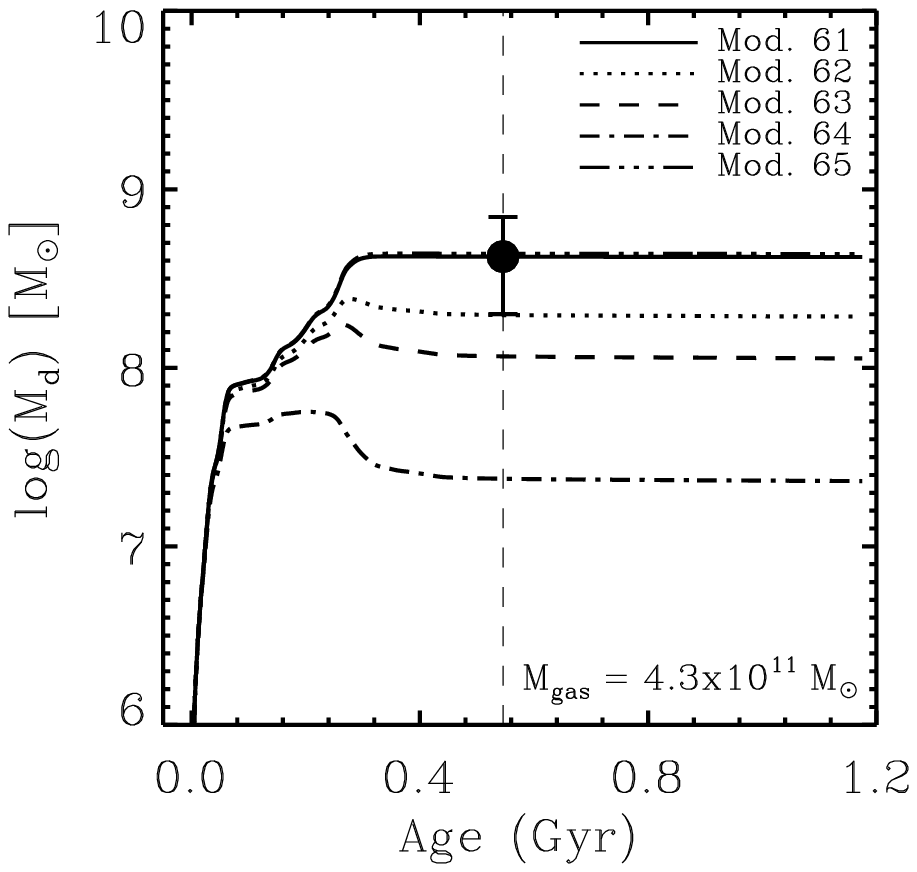}
  \includegraphics{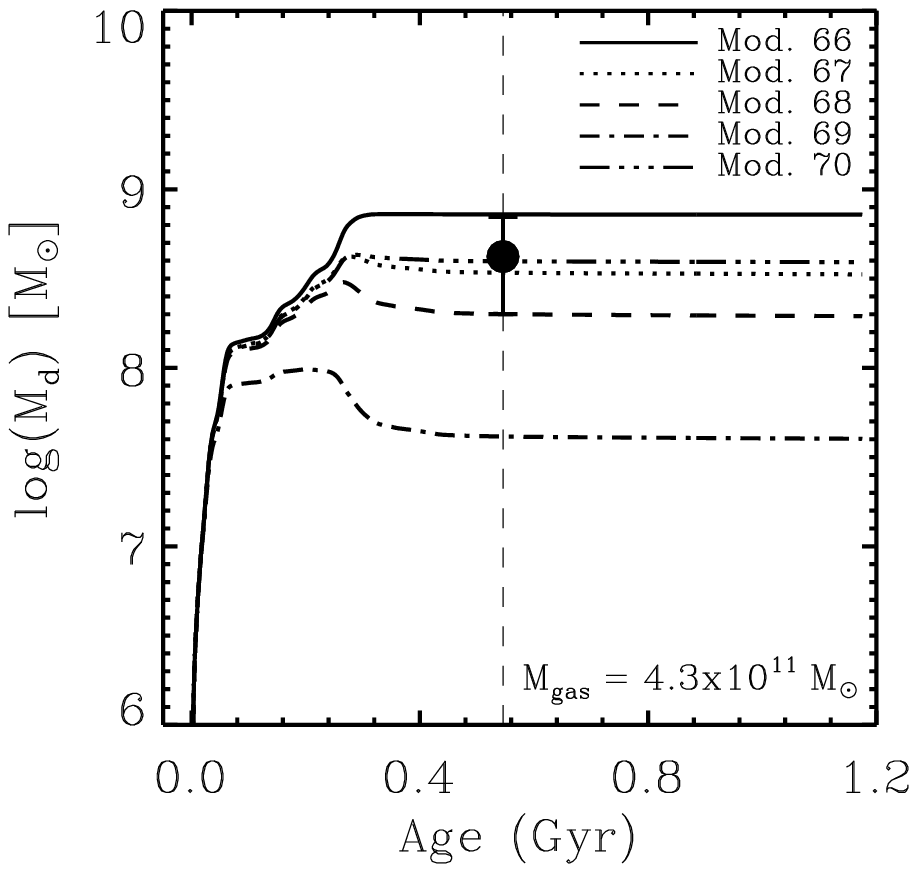}
  \includegraphics{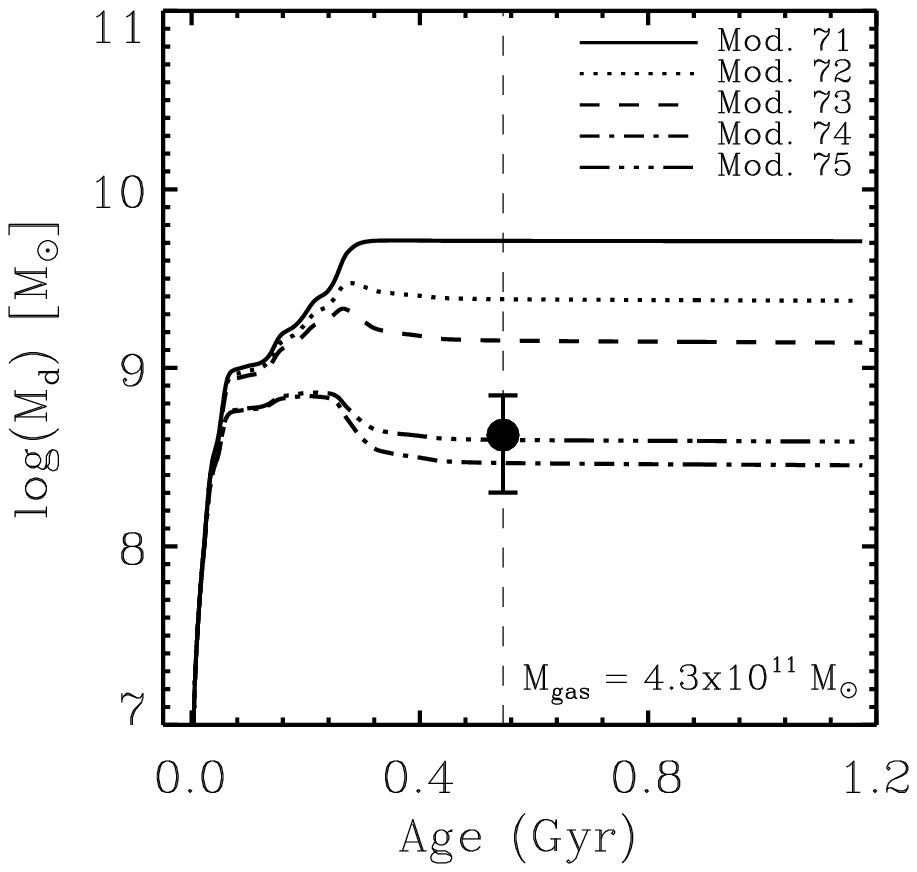}
  }
  \resizebox{15.9cm}{!}{
  \includegraphics{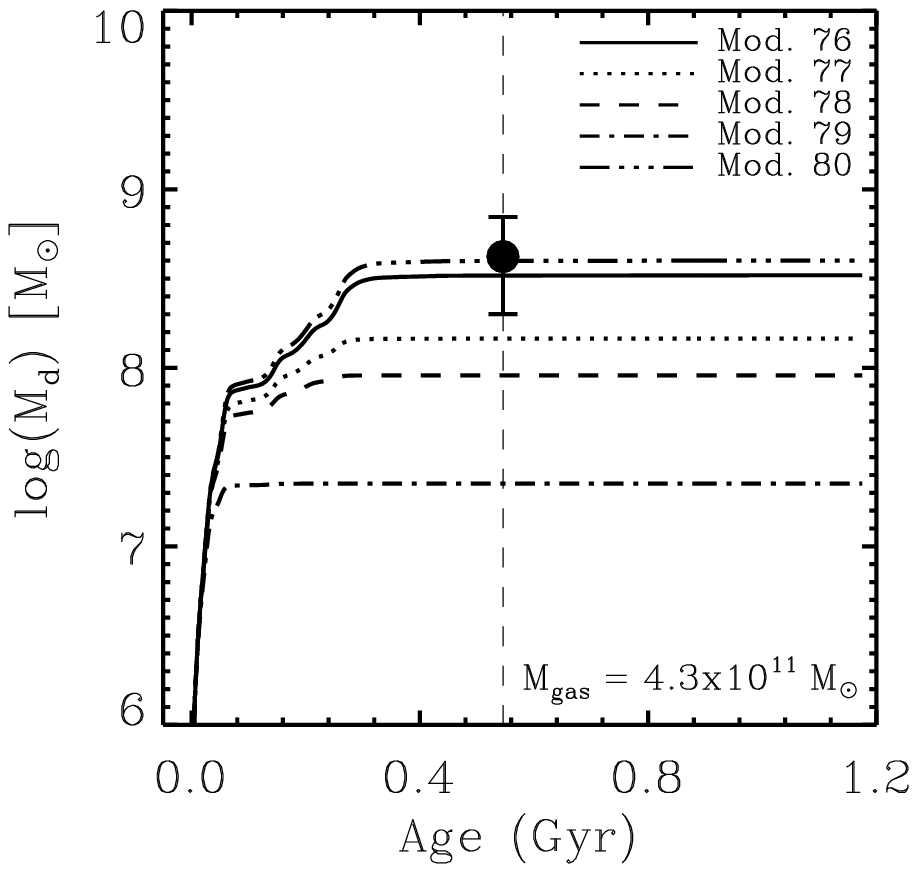}
  \includegraphics{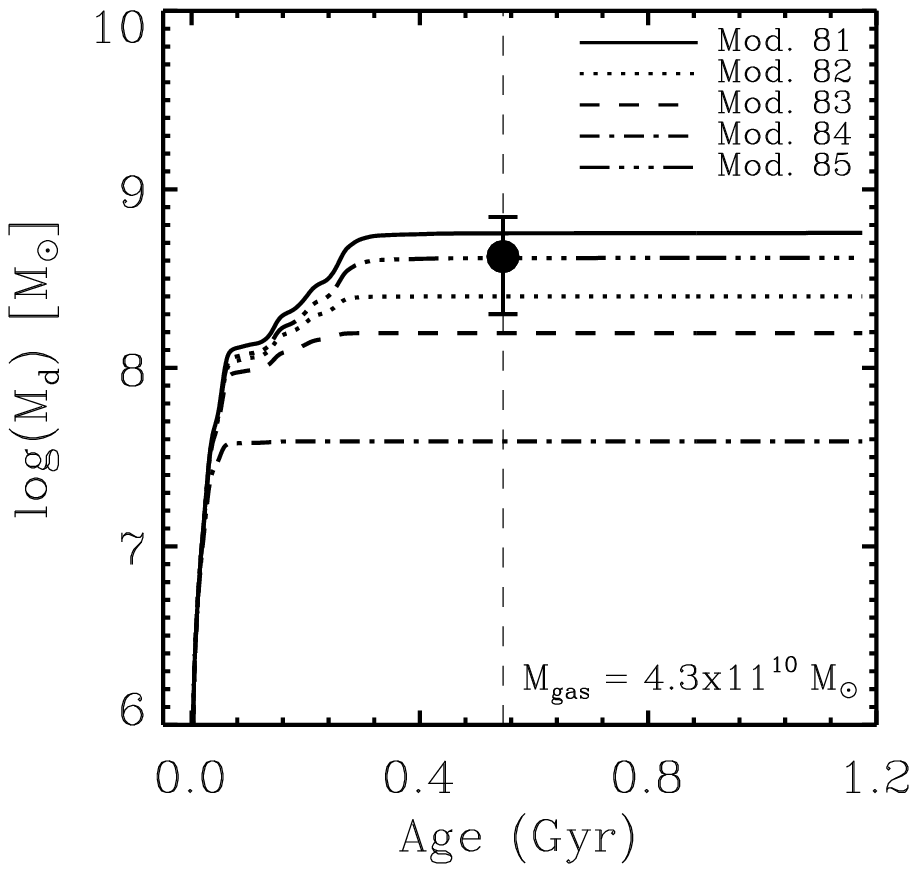}
  \includegraphics{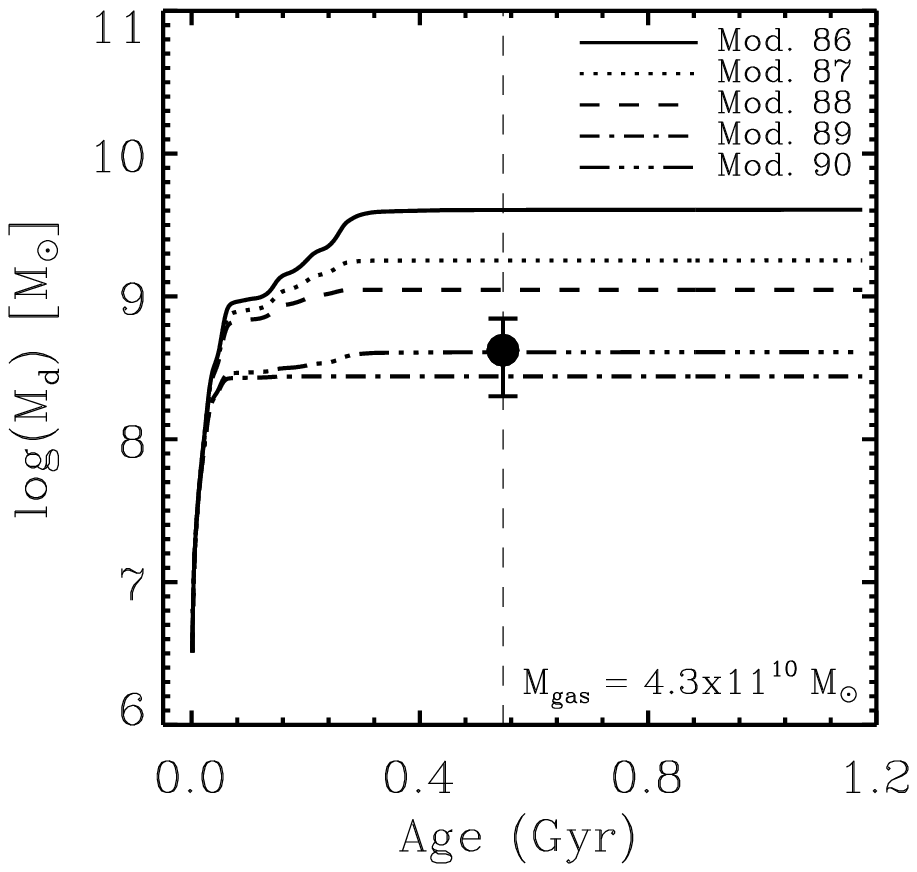}
  }
  \caption{Same as Fig. \ref{quasar}, but for the models with a normal IMF and a total mass of $1.3\cdot 10^{12}M_\odot$ \citep{Li07, Valiante09}.
          \label{quasar_g}}
  \end{figure*}

      \begin{table*}
  \begin{center}
  \caption{\label{parameters_s} Parameter values for models with a normal IMF and dynamically constrained total mass. Masses are given for $z = 6.42$. }
  \begin{tabular}{lllllllllll}

  Model & $\alpha$ & $\epsilon$ & $y_{\rm d}$ & $y_Z$ & $m_{\rm ISM}$ [$M_\odot$] & $\delta_{\rm ISM}$ & $\nu$ & $M_{\rm gas}$ [$M_\odot$] & $M_{\rm tot}$ [$M_\odot$] & Description \\
  \hline
  91     & 0.63     & -          & $8.81\cdot 10^{-4}$ & $3.00\cdot 10^{-2}$ & 0             & 0.00               & -     & $1.6\cdot 10^{10}$ & $4.5\cdot 10^{10}$ & Closed box, no dust destruction.\\
  92     & 0.63     & -          & $8.81\cdot 10^{-4}$ & $3.00\cdot 10^{-2}$ & 100           & 1.00               & 1.59  & $1.6\cdot 10^{10}$ & $4.5\cdot 10^{10}$ & Closed box.\\
  93     & 0.63     & -          & $8.81\cdot 10^{-4}$ & $3.00\cdot 10^{-2}$ & 200           & 2.00               & 3.17  & $1.6\cdot 10^{10}$ & $4.5\cdot 10^{10}$ & Closed box.\\
  94     & 0.63     & -          & $8.81\cdot 10^{-4}$ & $3.00\cdot 10^{-2}$ & 1000          & 10.0               & 15.9  & $1.6\cdot 10^{10}$ & $4.5\cdot 10^{10}$ & Closed box.\\
  95     & 0.63     & 1.50       & $8.81\cdot 10^{-4}$ & $3.00\cdot 10^{-2}$ & 0             & 0.00               & -     & $1.6\cdot 10^{10}$ & $4.5\cdot 10^{10}$ & Closed box, 'secondary dust'.\\[1mm]

  96     & 0.63     & -          & $1.52\cdot 10^{-3}$ & $3.00\cdot 10^{-2}$ & 0             & 0.00               & -     & $1.6\cdot 10^{10}$ & $4.5\cdot 10^{10}$ & Closed box, no dust destruction.\\
  97     & 0.63     & -          & $1.52\cdot 10^{-3}$ & $3.00\cdot 10^{-2}$ & 100           & 1.00               & 1.59  & $1.6\cdot 10^{10}$ & $4.5\cdot 10^{10}$ & Closed box.\\
  98     & 0.63     & -          & $1.52\cdot 10^{-3}$ & $3.00\cdot 10^{-2}$ & 200           & 2.00               & 3.17  & $1.6\cdot 10^{10}$ & $4.5\cdot 10^{10}$ & Closed box.\\
  99     & 0.63     & -          & $1.52\cdot 10^{-3}$ & $3.00\cdot 10^{-2}$ & 1000          & 10.0               & 15.9  & $1.6\cdot 10^{10}$ & $4.5\cdot 10^{10}$ & Closed box.\\
  100    & 0.63     & 1.50       & $1.52\cdot 10^{-3}$ & $3.00\cdot 10^{-2}$ & 0             & 0.00               & -     & $1.6\cdot 10^{10}$ & $4.5\cdot 10^{10}$ & Closed box, 'secondary dust'.\\[1mm]

  101    & 0.63     & -          & $1.08\cdot 10^{-2}$ & $3.00\cdot 10^{-2}$ & 0             & 0.00               & -     & $1.6\cdot 10^{10}$ & $4.5\cdot 10^{10}$ & Closed box, no dust destruction.\\
  102    & 0.63     & -          & $1.08\cdot 10^{-2}$ & $3.00\cdot 10^{-2}$ & 100           & 1.00               & 1.59  & $1.6\cdot 10^{10}$ & $4.5\cdot 10^{10}$ & Closed box.\\
  103    & 0.63     & -          & $1.08\cdot 10^{-2}$ & $3.00\cdot 10^{-2}$ & 200           & 2.00               & 3.17  & $1.6\cdot 10^{10}$ & $4.5\cdot 10^{10}$ & Closed box.\\
  104    & 0.63     & -          & $1.08\cdot 10^{-2}$ & $3.00\cdot 10^{-2}$ & 1000          & 10.0               & 15.9  & $1.6\cdot 10^{10}$ & $4.5\cdot 10^{10}$ & Closed box.\\
  105    & 0.63     & 1.00       & $1.08\cdot 10^{-2}$ & $3.00\cdot 10^{-2}$ & 0             & 0.00               & -     & $1.6\cdot 10^{10}$ & $4.5\cdot 10^{10}$ & Closed box, 'secondary dust'.\\[1mm]

  106    & 0.63     & -          & $8.81\cdot 10^{-4}$ & $3.00\cdot 10^{-2}$ & 0             & 0.00               & -     & $1.6\cdot 10^{10}$ & $4.5\cdot 10^{10}$ & Infall, no dust destruction.\\
  107    & 0.63     & -          & $8.81\cdot 10^{-4}$ & $3.00\cdot 10^{-2}$ & 100           & 1.00               & 1.59  & $1.6\cdot 10^{10}$ & $4.5\cdot 10^{10}$ & Infall\\
  108    & 0.63     & -          & $8.81\cdot 10^{-4}$ & $3.00\cdot 10^{-2}$ & 200           & 2.00               & 3.17  & $1.6\cdot 10^{10}$ & $4.5\cdot 10^{10}$ & Infall\\
  109    & 0.63     & -          & $8.81\cdot 10^{-4}$ & $3.00\cdot 10^{-2}$ & 1000          & 10.0               & 15.9  & $1.6\cdot 10^{10}$ & $4.5\cdot 10^{10}$ & Infall\\
  110    & 0.63     & {\it 0.97} & $8.81\cdot 10^{-4}$ & $3.00\cdot 10^{-2}$ & 0             & 0.00               & -     & $1.6\cdot 10^{10}$ & $4.5\cdot 10^{10}$ & Infall, 'sec. dust', max. $\epsilon$.\\[1mm]

  111    & 0.63     & -          & $1.52\cdot 10^{-3}$ & $3.00\cdot 10^{-2}$ & 0             & 0.00               & -     & $1.6\cdot 10^{10}$ & $4.5\cdot 10^{10}$ & Infall, no dust destruction.\\
  112    & 0.63     & -          & $1.52\cdot 10^{-3}$ & $3.00\cdot 10^{-2}$ & 100           & 1.00               & 1.59  & $1.6\cdot 10^{10}$ & $4.5\cdot 10^{10}$ & Infall\\
  113    & 0.63     & -          & $1.52\cdot 10^{-3}$ & $3.00\cdot 10^{-2}$ & 200           & 2.00               & 3.17  & $1.6\cdot 10^{10}$ & $4.5\cdot 10^{10}$ & Infall\\
  114    & 0.63     & -          & $1.52\cdot 10^{-3}$ & $3.00\cdot 10^{-2}$ & 1000          & 10.0               & 15.9  & $1.6\cdot 10^{10}$ & $4.5\cdot 10^{10}$ & Infall\\
  115    & 0.63     & {\it 0.95} & $1.52\cdot 10^{-3}$ & $3.00\cdot 10^{-2}$ & 0             & 0.00               & -     & $1.6\cdot 10^{10}$ & $4.5\cdot 10^{10}$ & Infall, 'sec. dust', max. $\epsilon$.\\[1mm]

  116    & 0.63     & -          & $1.08\cdot 10^{-2}$ & $3.00\cdot 10^{-2}$ & 0             & 0.00               & -     & $1.6\cdot 10^{10}$ & $4.5\cdot 10^{10}$ & Infall, no dust destruction.\\
  117    & 0.63     & -          & $1.08\cdot 10^{-2}$ & $3.00\cdot 10^{-2}$ & 100           & 1.00               & 1.59  & $1.6\cdot 10^{10}$ & $4.5\cdot 10^{10}$ & Infall\\
  118    & 0.63     & -          & $1.08\cdot 10^{-2}$ & $3.00\cdot 10^{-2}$ & 200           & 2.00               & 3.17  & $1.6\cdot 10^{10}$ & $4.5\cdot 10^{10}$ & Infall\\
  119    & 0.63     & -          & $1.08\cdot 10^{-2}$ & $3.00\cdot 10^{-2}$ & 1000          & 10.0               & 15.9  & $1.6\cdot 10^{10}$ & $4.5\cdot 10^{10}$ & Infall\\
  120    & 0.63     & {\it 0.64} & $1.08\cdot 10^{-2}$ & $3.00\cdot 10^{-2}$ & 0             & 0.00               & -     & $1.6\cdot 10^{10}$ & $4.5\cdot 10^{10}$ & Infall, 'sec. dust', max. $\epsilon$.\\
  \hline
  \end{tabular}
  \end{center}
  \end{table*}

  \begin{figure*}

  \resizebox{15.9cm}{!}{
  \includegraphics{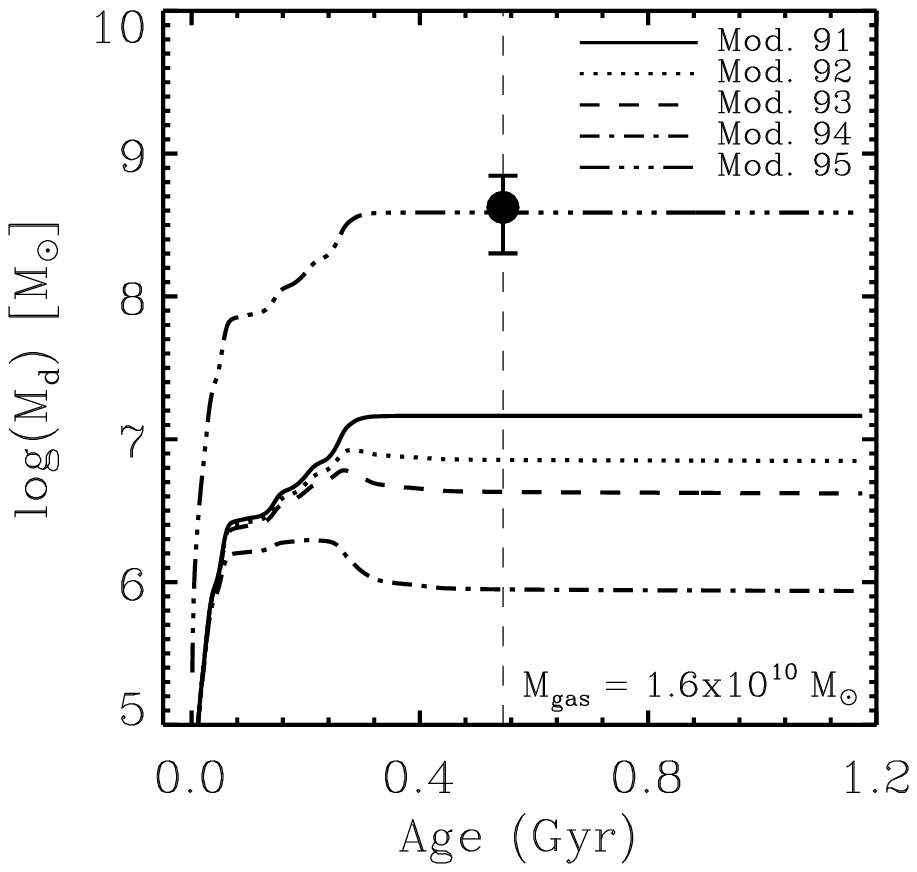}
  \includegraphics{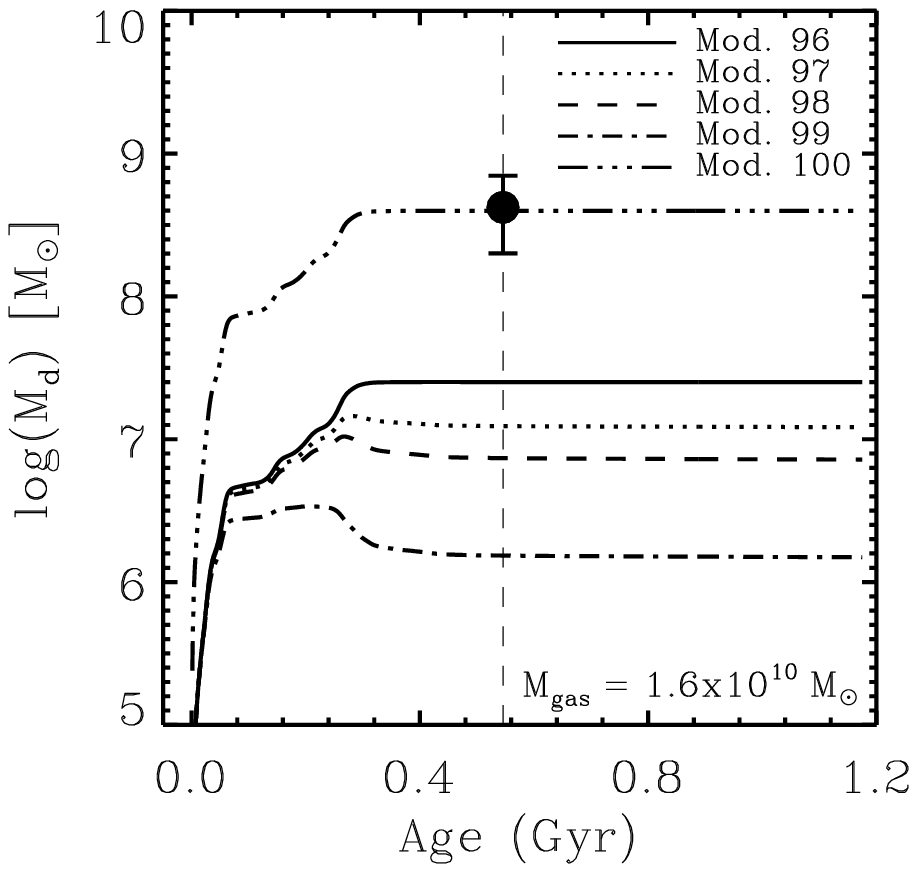}
  \includegraphics{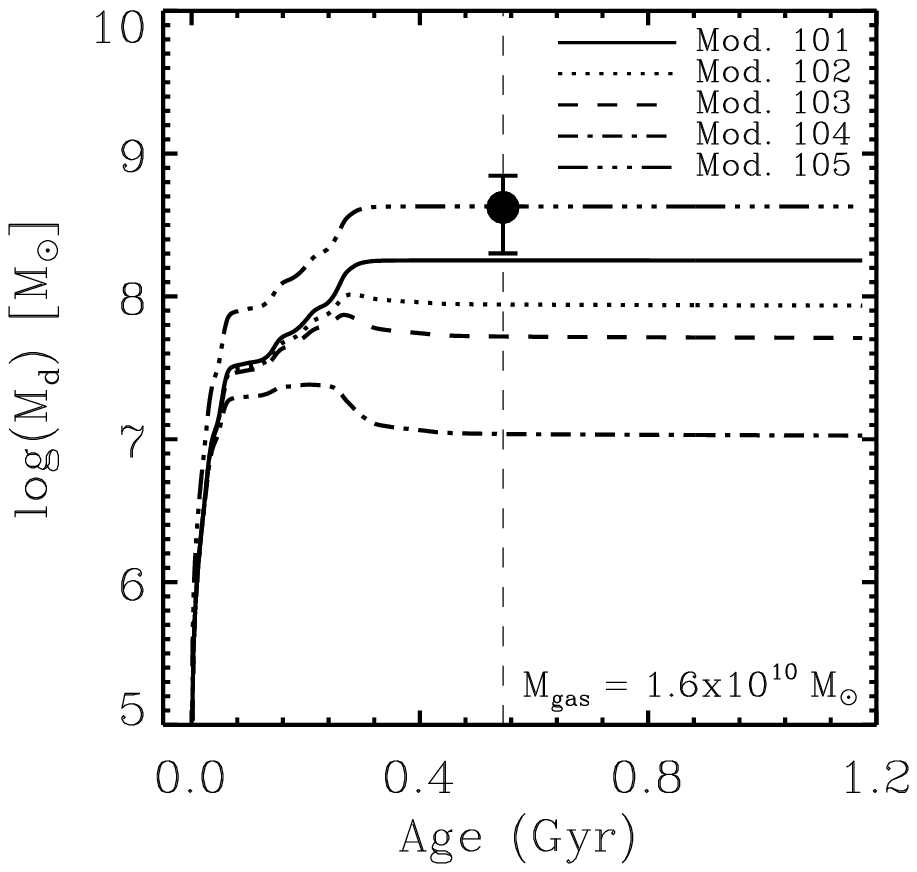}
  }
  \resizebox{15.9cm}{!}{
  \includegraphics{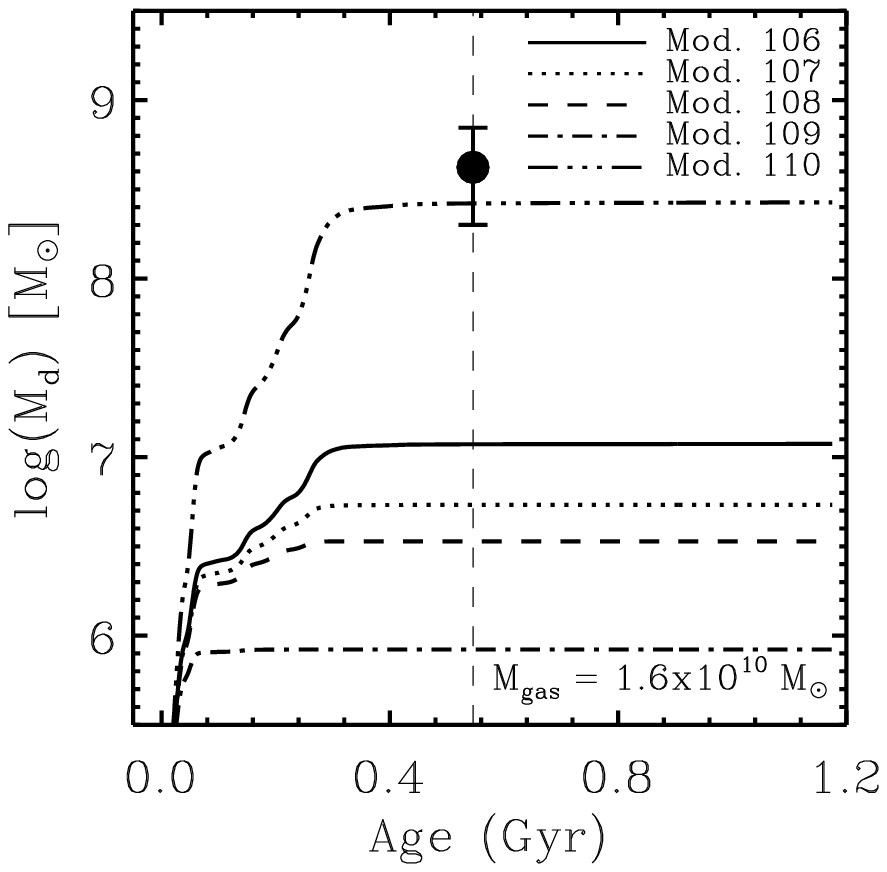}
  \includegraphics{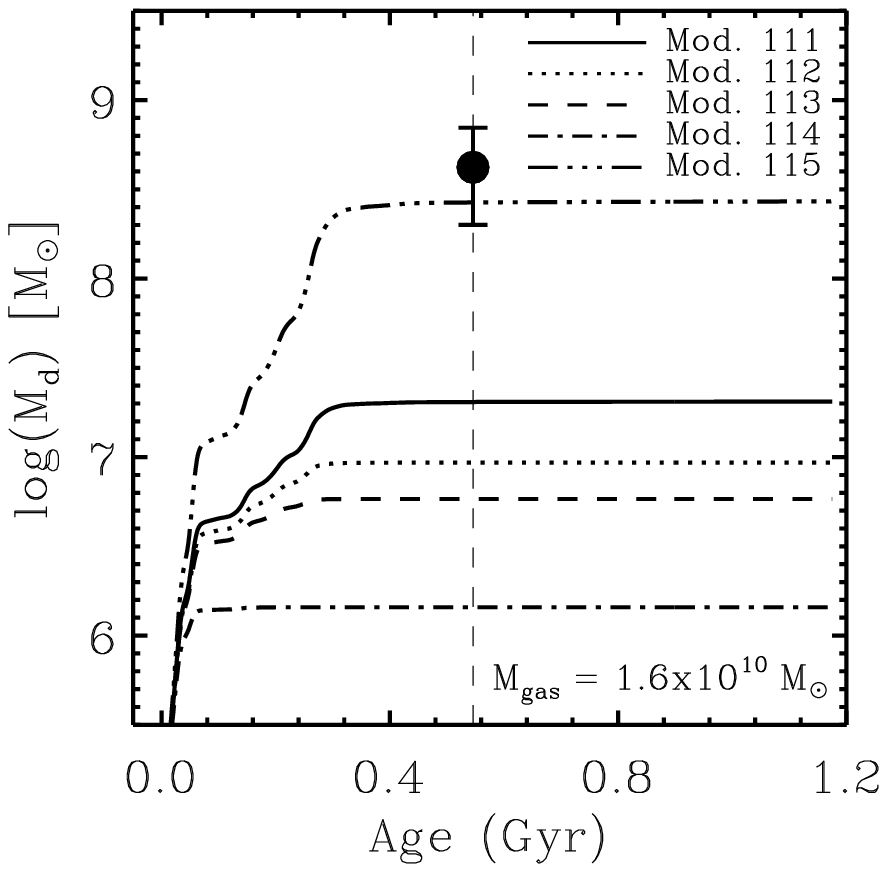}
  \includegraphics{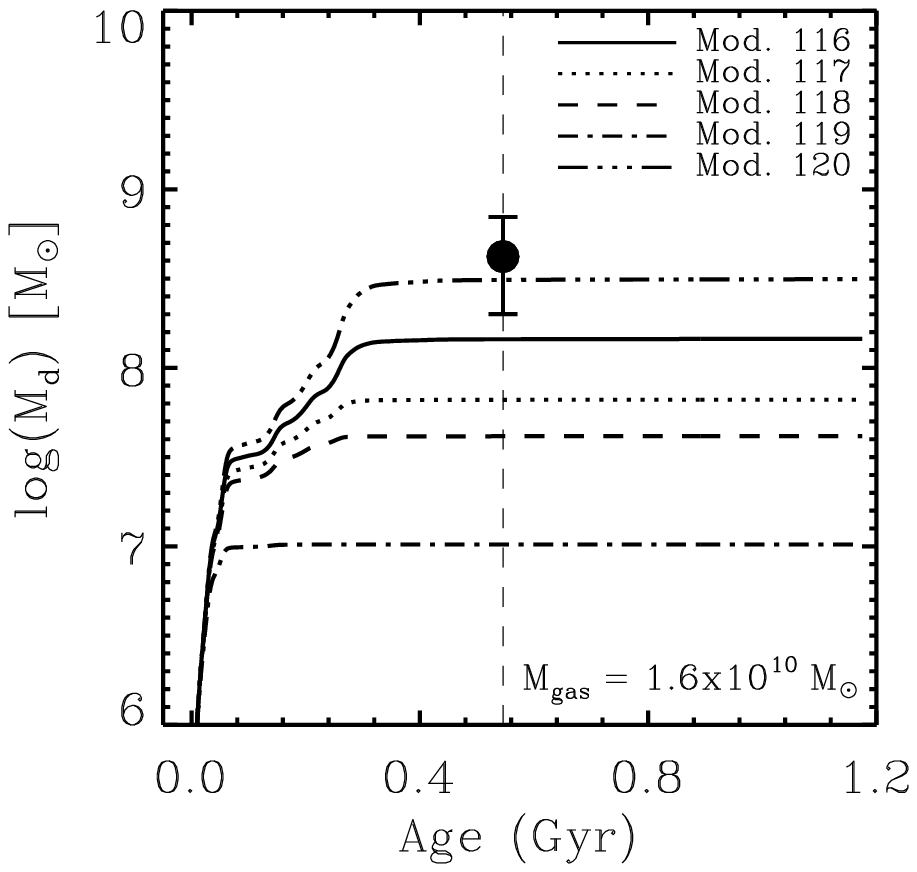}
  }
  \caption{Same as Fig. \ref{quasar}, but for the models with a normal IMF and a total mass of $4.5\cdot 10^{10}M_\odot$ as obtained from dynamical estimates
  \citep{Walter04}.
          \label{quasar_gs}}
  \end{figure*}

        \begin{table*}
  \begin{center}
  \caption{\label{parameters_sth} Parameter values for models with a top-heavy IMF and dynamically constrained total mass. Masses are given for $z = 6.42$.}
  \begin{tabular}{lllllllllll}

  Model & $\alpha$ & $\epsilon$ & $y_{\rm d}$ & $y_Z$ & $m_{\rm ISM}$ [$M_\odot$] & $\delta_{\rm ISM}$ & $\nu$ & $M_{\rm gas}$ [$M_\odot$] & $M_{\rm tot}$ [$M_\odot$] & Description \\
  \hline
  121    & 0.10     & -          & $8.81\cdot 10^{-4}$ & $6.50\cdot 10^{-1}$ & 0             & 0.00               & -     & $1.6\cdot 10^{10}$ & $4.5\cdot 10^{10}$ & Closed box, no dust destruction.\\
  122    & 0.10     & -          & $8.81\cdot 10^{-4}$ & $6.50\cdot 10^{-1}$ & 20            & 1.32               & 1.59  & $1.6\cdot 10^{10}$ & $4.5\cdot 10^{10}$ & Closed box.\\
  123    & 0.10     & -          & $8.81\cdot 10^{-4}$ & $6.50\cdot 10^{-1}$ & 100           & 6.60               & 3.17  & $1.6\cdot 10^{10}$ & $4.5\cdot 10^{10}$ & Closed box.\\
  124    & 0.10     & -          & $8.81\cdot 10^{-4}$ & $6.50\cdot 10^{-1}$ & 200           & 13.2               & 15.9  & $1.6\cdot 10^{10}$ & $4.5\cdot 10^{10}$ & Closed box.\\
  125    & 0.10     & 0.55       & $8.81\cdot 10^{-4}$ & $6.50\cdot 10^{-1}$ & 20            & 1.32               & -     & $1.6\cdot 10^{10}$ & $4.5\cdot 10^{10}$ & Closed box, 'secondary dust'.\\[1mm]
                                             
  126    & 0.10     & -          & $1.52\cdot 10^{-3}$ & $6.50\cdot 10^{-1}$ & 0             & 0.00               & -     & $1.6\cdot 10^{10}$ & $4.5\cdot 10^{10}$ & Closed box, no dust destruction.\\
  127    & 0.10     & -          & $1.52\cdot 10^{-3}$ & $6.50\cdot 10^{-1}$ & 20            & 1.32               & 1.59  & $1.6\cdot 10^{10}$ & $4.5\cdot 10^{10}$ & Closed box.\\
  128    & 0.10     & -          & $1.52\cdot 10^{-3}$ & $6.50\cdot 10^{-1}$ & 100           & 6.60               & 3.17  & $1.6\cdot 10^{10}$ & $4.5\cdot 10^{10}$ & Closed box.\\
  129    & 0.10     & -          & $1.52\cdot 10^{-3}$ & $6.50\cdot 10^{-1}$ & 200           & 13.2               & 15.9  & $1.6\cdot 10^{10}$ & $4.5\cdot 10^{10}$ & Closed box.\\
  130    & 0.10     & 0.50       & $1.52\cdot 10^{-3}$ & $6.50\cdot 10^{-1}$ & 20            & 1.32               & -     & $1.6\cdot 10^{10}$ & $4.5\cdot 10^{10}$ & Closed box, 'secondary dust'.\\[1mm]
                                             
  131    & 0.10     & -          & $1.08\cdot 10^{-2}$ & $6.50\cdot 10^{-1}$ & 0             & 0.00               & -     & $1.6\cdot 10^{10}$ & $4.5\cdot 10^{10}$ & Closed box, no dust destruction.\\
  132    & 0.10     & -          & $1.08\cdot 10^{-2}$ & $6.50\cdot 10^{-1}$ & 20            & 1.32               & 1.59  & $1.6\cdot 10^{10}$ & $4.5\cdot 10^{10}$ & Closed box.\\
  133    & 0.10     & -          & $1.08\cdot 10^{-2}$ & $6.50\cdot 10^{-1}$ & 100           & 6.60               & 3.17  & $1.6\cdot 10^{10}$ & $4.5\cdot 10^{10}$ & Closed box.\\
  134    & 0.10     & -          & $1.08\cdot 10^{-2}$ & $6.50\cdot 10^{-1}$ & 200           & 13.2               & 15.9  & $1.6\cdot 10^{10}$ & $4.5\cdot 10^{10}$ & Closed box.\\
  135    & 0.10     & 0.25       & $1.08\cdot 10^{-2}$ & $6.50\cdot 10^{-1}$ & 20            & 1.32               & 1.59  & $1.6\cdot 10^{10}$ & $4.5\cdot 10^{10}$ & Closed box, 'secondary dust'.\\[1mm]
                                             
  136    & 0.10     & -          & $8.81\cdot 10^{-4}$ & $6.50\cdot 10^{-1}$ & 0             & 0.00               & -     & $1.6\cdot 10^{10}$ & $4.5\cdot 10^{10}$ & Infall, no dust destruction.\\
  137    & 0.10     & -          & $8.81\cdot 10^{-4}$ & $6.50\cdot 10^{-1}$ & 20            & 1.32               & 1.59  & $1.6\cdot 10^{10}$ & $4.5\cdot 10^{10}$ & Infall\\
  138    & 0.10     & -          & $8.81\cdot 10^{-4}$ & $6.50\cdot 10^{-1}$ & 100           & 6.60               & 3.17  & $1.6\cdot 10^{10}$ & $4.5\cdot 10^{10}$ & Infall\\
  139    & 0.10     & -          & $8.81\cdot 10^{-4}$ & $6.50\cdot 10^{-1}$ & 200           & 13.2               & 15.9  & $1.6\cdot 10^{10}$ & $4.5\cdot 10^{10}$ & Infall\\
  140    & 0.10     & 0.70       & $8.81\cdot 10^{-4}$ & $6.50\cdot 10^{-1}$ & 20            & 1.32               & -     & $1.6\cdot 10^{10}$ & $4.5\cdot 10^{10}$ & Infall, 'secondary dust'.\\[1mm]
                                             
  141    & 0.10     & -          & $1.52\cdot 10^{-3}$ & $6.50\cdot 10^{-1}$ & 0             & 0.00               & -     & $1.6\cdot 10^{10}$ & $4.5\cdot 10^{10}$ & Infall, no dust destruction.\\
  142    & 0.10     & -          & $1.52\cdot 10^{-3}$ & $6.50\cdot 10^{-1}$ & 20            & 1.32               & 1.59  & $1.6\cdot 10^{10}$ & $4.5\cdot 10^{10}$ & Infall\\
  143    & 0.10     & -          & $1.52\cdot 10^{-3}$ & $6.50\cdot 10^{-1}$ & 100           & 6.60               & 3.17  & $1.6\cdot 10^{10}$ & $4.5\cdot 10^{10}$ & Infall\\
  144    & 0.10     & -          & $1.52\cdot 10^{-3}$ & $6.50\cdot 10^{-1}$ & 200           & 13.2               & 15.9  & $1.6\cdot 10^{10}$ & $4.5\cdot 10^{10}$ & Infall\\
  145    & 0.10     & 0.65       & $1.52\cdot 10^{-3}$ & $6.50\cdot 10^{-1}$ & 20            & 1.32               & -     & $1.6\cdot 10^{10}$ & $4.5\cdot 10^{10}$ & Infall, 'secondary dust'.\\[1mm]
                                             
  146    & 0.10     & -          & $1.08\cdot 10^{-2}$ & $6.50\cdot 10^{-1}$ & 0             & 0.00               & -     & $1.6\cdot 10^{10}$ & $4.5\cdot 10^{10}$ & Infall, no dust destruction.\\
  147    & 0.10     & -          & $1.08\cdot 10^{-2}$ & $6.50\cdot 10^{-1}$ & 20            & 1.32               & 1.59  & $1.6\cdot 10^{10}$ & $4.5\cdot 10^{10}$ & Infall\\
  148    & 0.10     & -          & $1.08\cdot 10^{-2}$ & $6.50\cdot 10^{-1}$ & 100           & 6.60               & 3.17  & $1.6\cdot 10^{10}$ & $4.5\cdot 10^{10}$ & Infall\\
  149    & 0.10     & -          & $1.08\cdot 10^{-2}$ & $6.50\cdot 10^{-1}$ & 200           & 13.2               & 15.9  & $1.6\cdot 10^{10}$ & $4.5\cdot 10^{10}$ & Infall\\
  150    & 0.10     & 0.30       & $1.08\cdot 10^{-2}$ & $6.50\cdot 10^{-1}$ & 20            & 1.32               & -     & $1.6\cdot 10^{10}$ & $4.5\cdot 10^{10}$ & Infall, 'secondary dust'.\\
  \hline
  \end{tabular}
  \end{center}
  \end{table*}

  \begin{figure*}

  \resizebox{15.9cm}{!}{
  \includegraphics{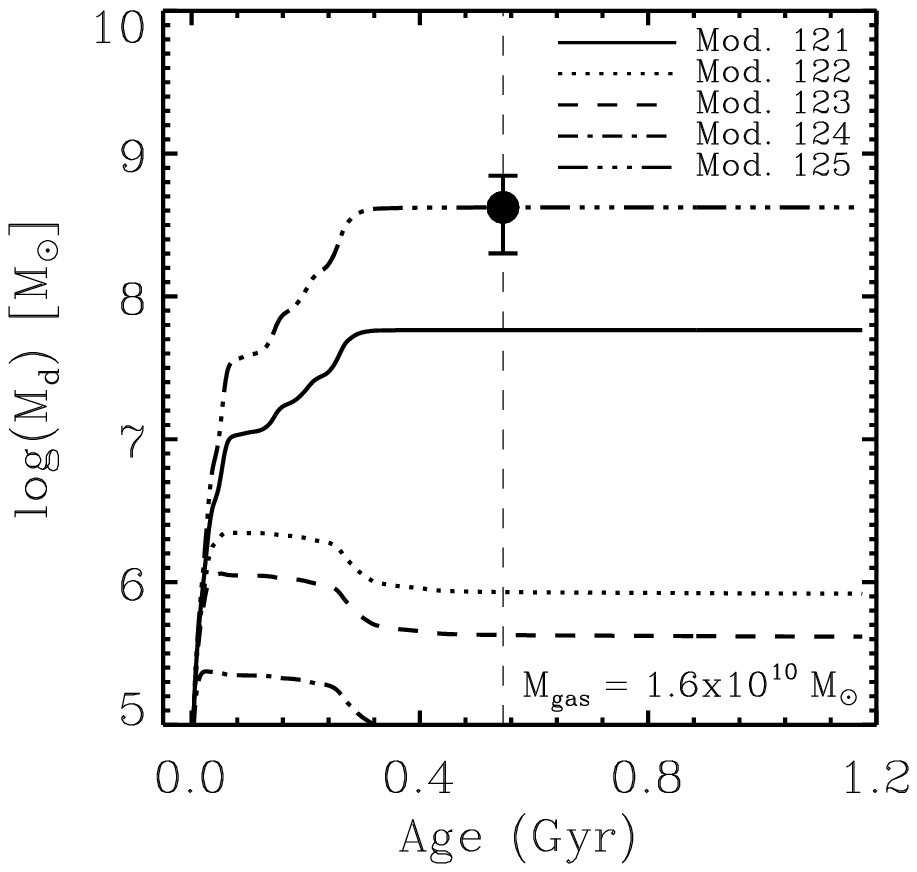}
  \includegraphics{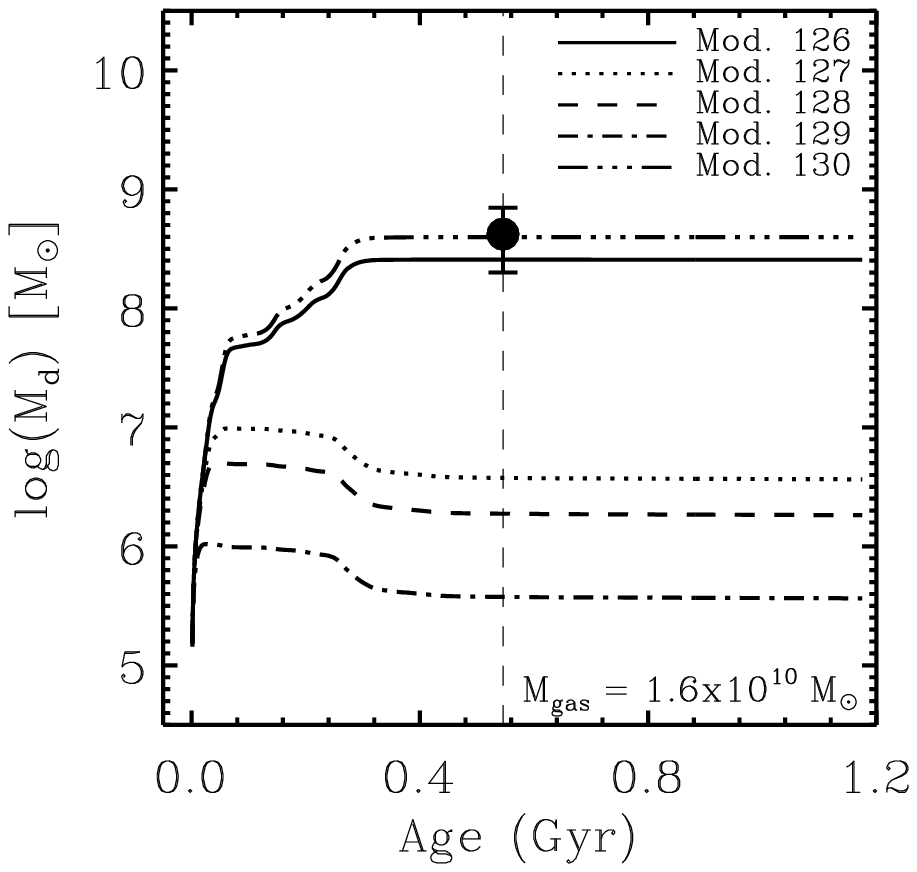}
  \includegraphics{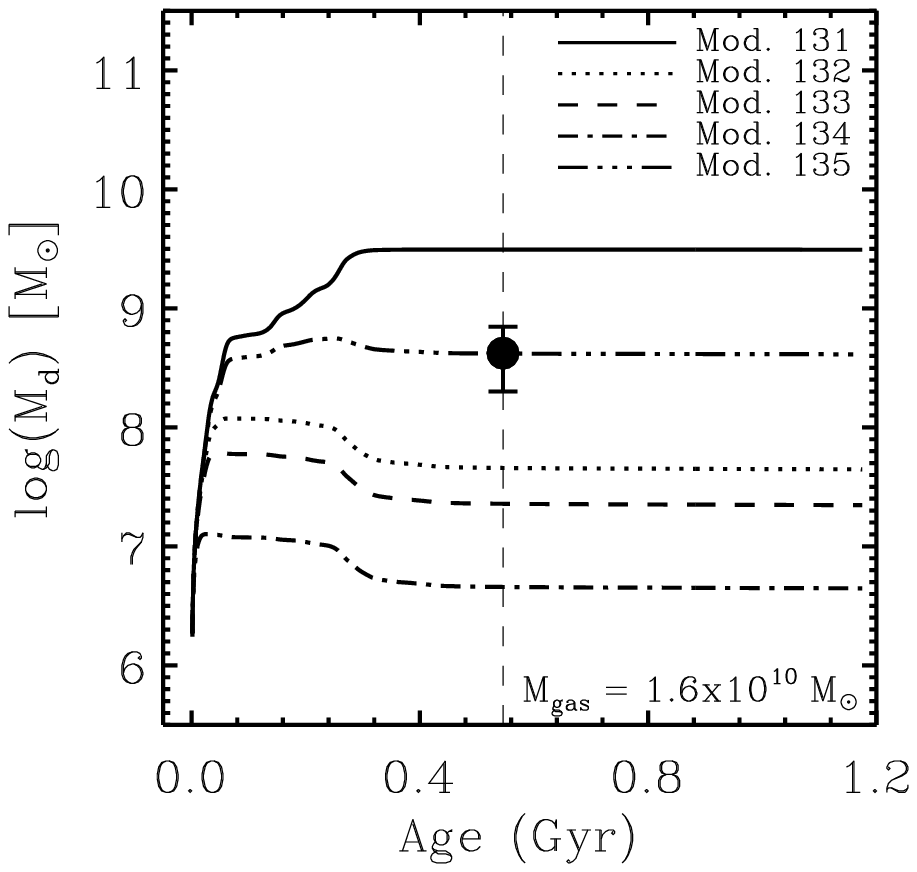}
  }
  \resizebox{15.9cm}{!}{
  \includegraphics{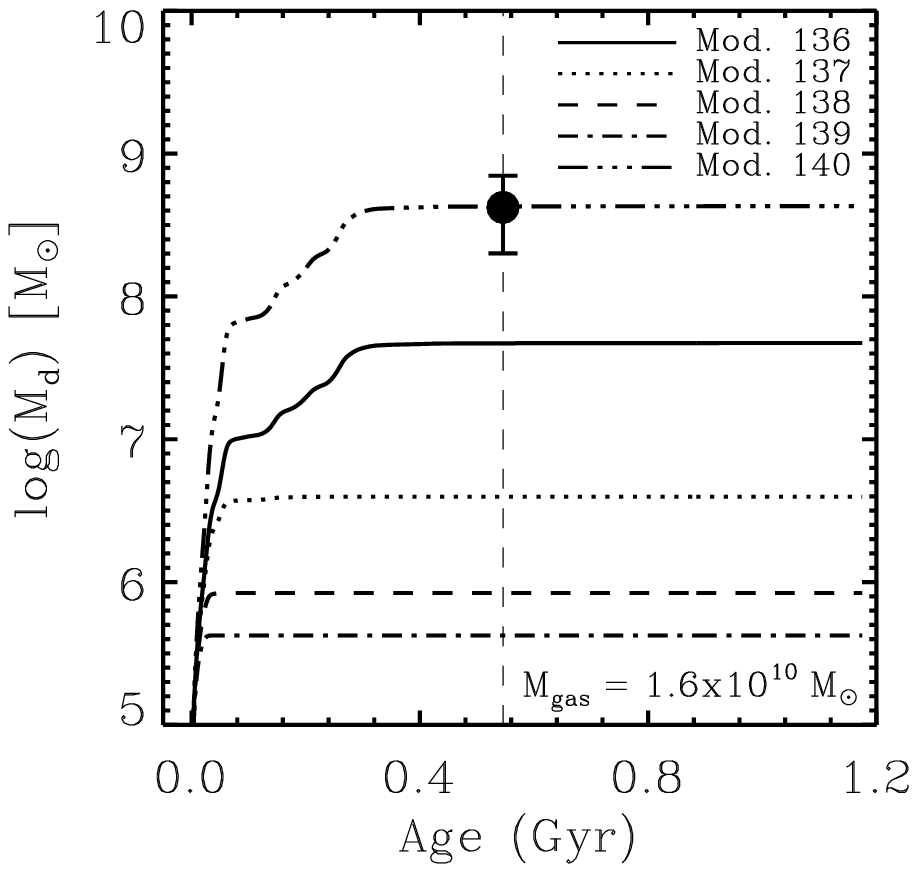}
  \includegraphics{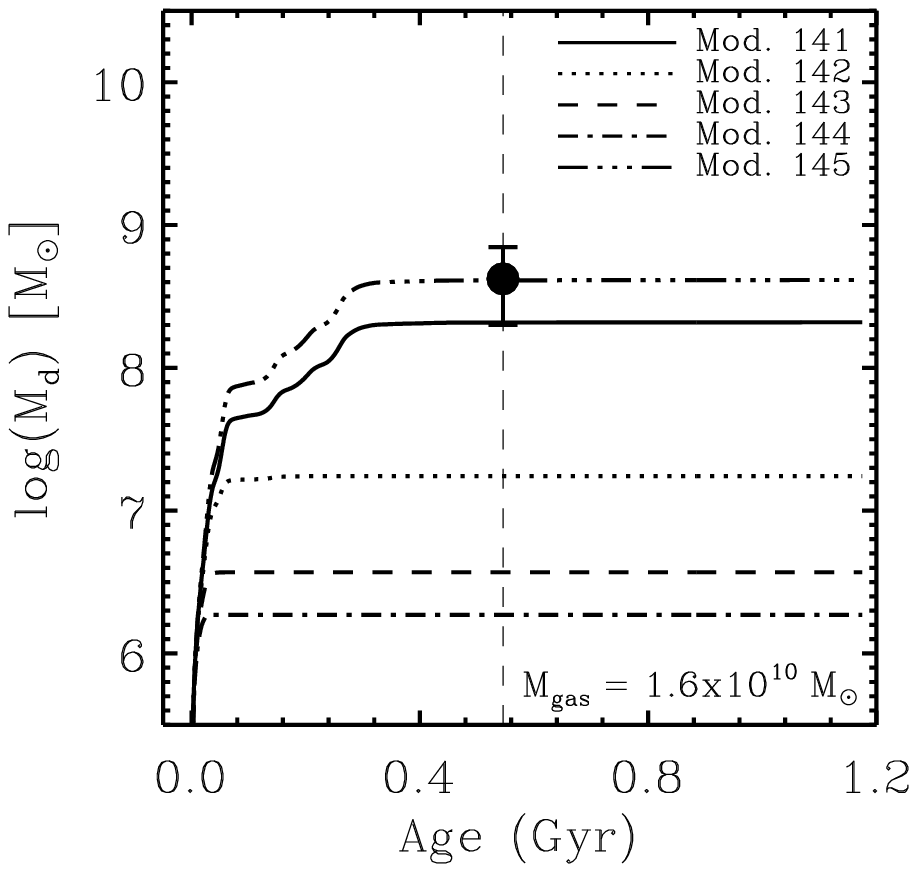}
  \includegraphics{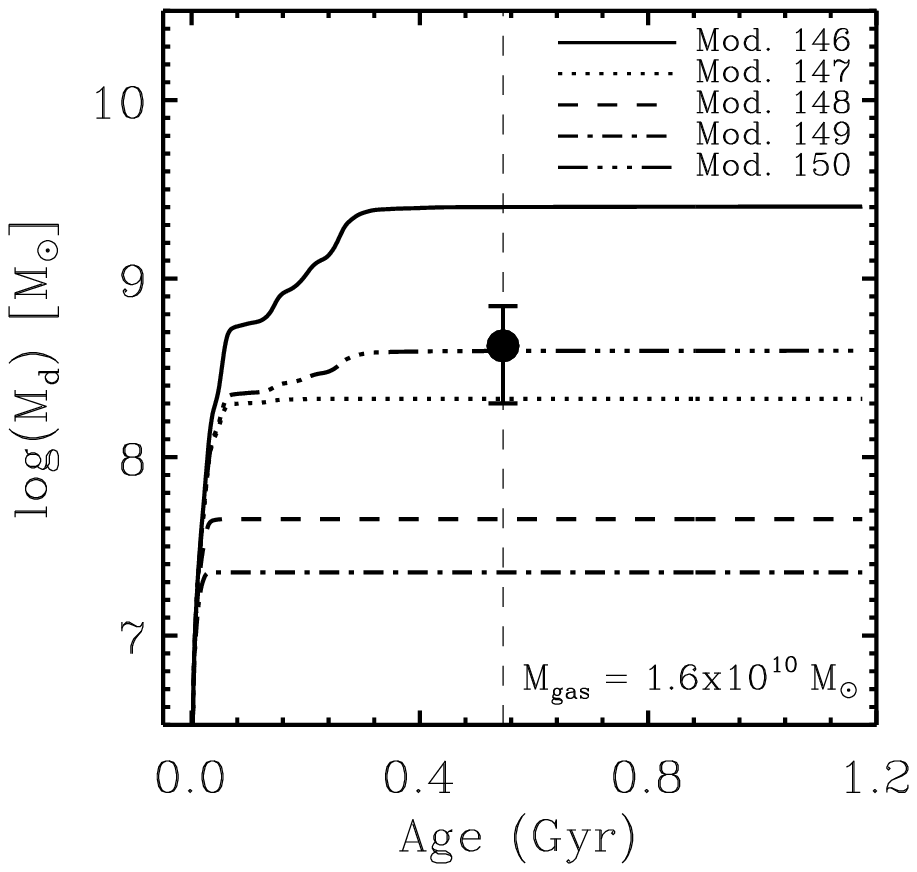}
  }
  \caption{Same as Fig. \ref{quasar}, but for the models with a top-heavy IMF and a total mass of $4.5\cdot 10^{10}M_\odot$ as obtained from dynamical estimates
  \citep{Walter04}.
          \label{quasar_sth}}
  \end{figure*}

\end{document}